\documentclass[11pt,a4paper]{article}
\pdfoutput=1
\usepackage{jcappub}
\usepackage{rotating}
\usepackage{graphicx}
\usepackage{epsfig}
\usepackage{amsmath}
\usepackage{amssymb}
\usepackage{subfigure}
\usepackage{time}
\usepackage{relsize}
\usepackage{xcolor}
\usepackage[utf8]{inputenc}

\title{Black holes with scalar hair in light of the Event Horizon Telescope}

\author[a]{Mohsen Khodadi,}
\author[a]{Alireza Allahyari,}
\author[b]{Sunny Vagnozzi,}
\author[c]{and David F. Mota}

\affiliation[a]{School of Astronomy, Institute for Research in Fundamental Sciences (IPM), P.O. Box 19395-5531, Tehran, Iran}
\affiliation[b]{Kavli Institute for Cosmology (KICC) and Institute of Astronomy, University of Cambridge, Madingley Road, Cambridge CB3 0HA, United Kingdom}
\affiliation[c]{Institute of Theoretical Astrophysics, University of Oslo, P.O. Box 1029 Blindern, N-0315 Oslo, Norway}

\emailAdd{m.khodadi@ipm.ir}
\emailAdd{alireza.al@ipm.ir}
\emailAdd{sunny.vagnozzi@ast.cam.ac.uk}
\emailAdd{d.f.mota@astro.uio.no}

\abstract{Searching for violations of the no-hair theorem (NHT) is a powerful way to test gravity, and more generally fundamental physics, particularly with regards to the existence of additional scalar fields. The first observation of a black hole (BH) shadow by the Event Horizon Telescope (EHT) has opened a new direct window onto tests of gravity in the strong-field regime, including probes of violations of the NHT. We consider two scenarios described by the Einstein-Maxwell equations of General Relativity and electromagnetism, to which we add a scalar field. In the first case we consider a minimally-coupled scalar field with a potential, whereas in the second case the field is conformally-coupled to curvature. In both scenarios we construct charged BH solutions, which are found to carry primary scalar hair. We then compute the shadows cast by these two BHs as a function of their electric charge and scalar hair parameter. Comparing these shadows to the shadow of M87* recently imaged by the EHT collaboration, we set constraints on the amount of scalar hair carried by these two BHs. The conformally-coupled case admits a regime for the hair parameter, compatible with EHT constraints, describing a so-called mutated Reissner-Nordstr\"{o}m BH: this solution was recently found to effectively mimic a wormhole. Our work provides novel constraints on fundamental physics, and in particular on violations of the no-hair theorem and the existence of additional scalar fields, from the shadow of M87*.}

\begin{document}
\maketitle
\flushbottom

\section{Introduction}
\label{sec:intro}

All black hole (BH) solutions of the coupled Einstein-Maxwell equations of General Relativity (GR) and electromagnetism (EM) can be described solely in terms of three classical observable parameters: mass $M$, spin $J$, and electric charge $Q$. To put it in the words of John Archibald Wheeler, ``\textit{black holes have no hair}'', where ``hair'' is loosely used to indicate all information regarding the matter which formed or fell into the BH, all of which is supposedly lost beyond the event horizon. This is the essence of the so-called \textit{no-hair theorem} (NHT), sometimes also referred to as uniqueness theorem: the theorem states that Kerr-Newman BHs are the only possible stationary, axisymmetric, regular, and asymptotically flat BH solutions of the vacuum Einstein-Maxwell equations~\cite{Israel:1967wq,Israel:1967za,Carter:1971zc}.

There is, at present, no sign of tension between the NHT and astrophysical observations. Yet, there are plenty of hints suggesting that our understanding of gravity might be incomplete. Some of these hints are somewhat more fundamental, including the quest for a quantum theory of gravity, or the conflict between Hawking BH radiation and unitary evolution in quantum mechanics, as exemplified by the BH information paradox~\cite{Hawking:1974sw}. Other hints are more observational in nature, including identifying the nature of the elusive dark matter (DM) and dark energy (DE), which drive structure formation and the late-time accelerated expansion of the Universe respectively. Modifications to Einstein's GR and/or additional fields beyond those of the Standard Model could address these problems, possibly providing the ingredients for explaining an early epoch of cosmic acceleration, dubbed inflation. Regardless of what turns out to be the explanation for DM, DE, or inflation, there is good reason to believe that in the presence of new physics, the NHT might be a valid approximation given current precision, but an approximation nonetheless.

This bring us to the question of whether the NHT can be evaded. A short answer is ``\textit{yes, but not without difficulty}''. In recent years, there have been several attempts to endow BHs with hair by means of additional matter fields. The metric of the resulting hairy BHs (HBHs) presents a permanent deformation, while the HBH horizon is maintained regular and its space-time asymptotically flat. In most of these constructions, the symmetries of the non-hairy metric (e.g. stationarity and axisymmetry) are preserved by the hairy metric, but not by the full (BH plus additional field) solution, as the additional field is usually required to have a non-trivial profile. Therefore, these solutions bypass the NHT by violating the basic requirement of stationarity and/or axisymmetry at the level of the full solution. However, as long as stationarity remains an isometry of the HBH, the resulting object can still play an important role in scenarios of astrophysical interest. For an inevitably incomplete list of important studies related to the construction and identification of observational signatures of HBHs in various theories of gravity or in the presence of various different types of matter fields, see for instance~\cite{Bronnikov:1973fh,Bronnikov:1977is,Mavromatos:1995kc,Kanti:1995vq,
Kleihaus:2000kg,Bronnikov:2005gm,Bronnikov:2006fu,Nadalini:2007qi,Rinaldi:2012vy,
Banerjee:2013yua,Brito:2013xaa,Babichev:2013cya,Sotiriou:2013qea,Herdeiro:2014goa,
Herdeiro:2014ima,Brihaye:2014nba,Herdeiro:2015gia,Sakalli:2015nza,Kleihaus:2015iea,
Cunha:2015yba,Herdeiro:2016tmi,Sanchis-Gual:2016tcm,Delgado:2016jxq,Cunha:2016bjh,Franchini:2016yvq,Herdeiro:2017oxy,
Heisenberg:2017xda,Herdeiro:2017phl,Erices:2017izj,Cisterna:2017jmv,
Antoniou:2017acq,Jusufi:2017drg,Tattersall:2018map,Brihaye:2018woc,Delgado:2018khf,
Herdeiro:2018daq,Lee:2018zym,Cisterna:2018mww,Raposo:2018xkf,Bernardo:2019mmx,
Filippini:2019cqk,Hong:2019mcj,Javed:2019rrg,VanAelst:2019kku,BenAchour:2019fdf,
Tattersall:2019nmh,Kim:2019hfp,Herdeiro:2020xmb,Mahapatra:2020wym,Hong:2020miv}.

One way to violate the NHT through additional fields is by means of scalar fields, \textit{i.e.} within a system described by the Einstein-Hilbert-Klein-Gordon equations. This is a possibility which is very interesting in its own right given how scalar fields, be they fundamental or effective/composite, are not only ubiquitous in nature (for instance, ultralight scalars are omnipresent within string theory~\cite{Svrcek:2006yi,Arvanitaki:2009fg,Visinelli:2018utg}), but are also arguably among the most economical and motivated candidates for explaining DM, DE, and inflation.~\footnote{For an inevitably incomplete list of examples to back up this statement, see e.g.~\cite{Linde:1983gd,Ratra:1987rm,Freese:1990rb,McDonald:1993ex,Peebles:1998qn,Bezrukov:2007ep,Chamseddine:2013kea,Rinaldi:2014yta,Myrzakulov:2015qaa,Rinaldi:2016oqp,Vagnozzi:2017ilo,Benisty:2018oyy,Caputo:2019joi,Benisty:2020nuu,DiLuzio:2020wdo} for examples of models where scalar fields play the role of DM, DE, or drive inflation, and/or unify one or more of these aspects, both in the context of General Relativity as well as within extensions thereof.} The main difficulty when trying to evade the NHT is to ensure that the scalar field remains regular at the BH horizon(s). One way of doing so is to explicitly introduce a new scale in the gravitational sector of the theory via a cosmological constant: that is, by considering BHs in asymptotically de Sitter (dS) or anti-de Sitter (AdS) space. Some of these attempts led to stable BHs with interesting physical properties, see e.g.~\cite{Dotti:2007cp,Kolyvaris:2009pc,Bardoux:2012tr}. A related possibility is that of still introducing a new scale, but this time in the scalar sector instead of in the gravitational sector~\cite{Charmousis:2009cm,Kolyvaris:2011fk}. An important question in the quest for controlled violations of the NHT through scalar fields is that of finding HBH solutions with primary scalar hair (\textit{i.e.} a conserved charge): most, albeit not all, known HBH solutions are instead characterized by secondary hair, which do not make it possible to continuously connect the HBH solution to a BH solution with the same mass but no scalar field. See~\cite{Herdeiro:2015waa} for a recent review on asymptotically flat BHs with scalar hair, and~\cite{Cardoso:2016ryw} for a recent review on observational tests of the NHT.

Besides the fundamental realization that our understanding of gravity is likely incomplete, there are at least two very valid reasons for why exploring controlled violations of the NHT is timely and interesting. The first reason is somewhat more theoretical, and is related to the AdS/CFT correspondence (or gauge/gravity duality)~\cite{Maldacena:1997re}. The AdS/CFT correspondence connects strongly coupled $d$-dimensional conformal field theories to weakly coupled $(d+1)$-dimensional gravitational theories, where the gravitational part of the system is described by an extremal charged BH in AdS space-time. Recent developments in our understanding of the gauge/gravity duality call for a deeper investigation of the behaviour of matter fields in the vicinity of charged BHs~\cite{Gubser:2005ih}, which might develop hair. A better understanding of hairy charged BHs will also likely open new windows into our understanding of a number of condensed matter systems which are especially interesting in light of the AdS/CFT correspondence. Examples include holographic superconductors (where the gravity sector contains a gauge field and a charged scalar field)~\cite{Horowitz:2009ij} and quantum liquids (many-body systems with finite $U(1)$ charge density at zero temperature)~\cite{Liu:2009dm}.

The second reason why it is timely to study controlled violations of the NHT is instead observational in nature. In the over 100 years since GR was developed we have been steadily accumulating indirect evidence for the existence of BHs, in a wide range of astrophysical environments and masses. At the same time, we have been unable to test that which is the very defining property of BHs, the existence of an event horizon (EH)~\cite{Einstein:1916vd,Schwarzschild:1916uq}. Two novel observational channels promise to test the geometry of space-time in the vicinity of the EH to exquisite accuracy: gravitational waves (GWs), and very-long-baseline-interferometry (VLBI) imaging of BH shadows. GW observations have already enabled unprecedented tests of GR~\cite{Abbott:2018lct} and ruled out, or severely constrained, many theoretically valid alternatives (see for instance~\cite{Creminelli:2017sry,Sakstein:2017xjx,Ezquiaga:2017ekz,Baker:2017hug,Boran:2017rdn,Arai:2017hxj,Amendola:2017orw,Visinelli:2017bny,Gumrukcuoglu:2017ijh,Cai:2018rzd,Pardo:2018ipy,Casalino:2018tcd,Ganz:2018vzg,Casalino:2018wnc,Odintsov:2019clh}). See e.g.~\cite{Barack:2018yly} for a recent review on future tests of fundamental physics from GWs and BHs.

VLBI imaging of BH shadows is instead an extremely novel probe, but at the same time a very promising one. A BH surrounded by a geometrically thick, optically thin emission region (e.g. an accretion disk) will appear to a distant observer as a central dark region, the so-called BH shadow, against the backdrop of a bright ring~\cite{Luminet:1979nyg,Cunha:2018acu}. The BH shadow is the apparent (\textit{i.e.} gravitationally lensed) image of the photon sphere, the region of space-time where photons travel along unstable orbits. In other words, the BH shadow is the closed curve on the sky separating capture orbits and scattering orbits. For a Schwarzschild BH of mass $M$, the radius of the BH shadow is $3\sqrt{3}M$, which is larger than both the event horizon and photon sphere radii ($2M$ and $3M$ respectively). Therefore, the BH shadow probes the geometry of space-time in the proximity of the event horizon. See e.g.~\cite{Dokuchaev:2019jqq} for a recent review on BH shadows.

In April 2019, the Event Horizon Telescope (EHT) collaboration, a VLBI radio telescope array operating at millimeter wavelengths with Earth-wide baseline, delivered what is probably one of the most iconic images of 21st century physics: the shadow of M87*, the supermassive BH residing at the center of the nearby supergiant elliptical galaxy Messier 87~\cite{Akiyama:2019cqa,Akiyama:2019brx,Akiyama:2019sww,Akiyama:2019bqs,Akiyama:2019fyp,Akiyama:2019eap}. Thanks to this extraordinary feat, VLBI BH shadow imaging finally became a reality. Early work had already established that VLBI BH shadow imaging can in principle be used to test violations of the NHT~\cite{Johannsen:2010ru,Johannsen:2015hib,Johannsen:2015mdd}. More generally, several works following the EHT detection have examined the possibility of constraining fundamental physics from the image of M87*'s shadow, mostly in the context of deviations from GR (some of which envisaging violations of the NHT). For an incomplete list of works in this direction, see for instance~\cite{Moffat:2019uxp,Giddings:2019jwy,Wei:2019pjf,Shaikh:2019fpu,Tamburini:2019vrf,Davoudiasl:2019nlo,Bambi:2019tjh,Konoplya:2019nzp,Contreras:2019nih,Bar:2019pnz,Jusufi:2019nrn,Vagnozzi:2019apd,Banerjee:2019cjk,Roy:2019esk,Long:2019nox,Zhu:2019ura,Contreras:2019cmf,Qi:2019zdk,Neves:2019lio,Pavlovic:2019rim,Biswas:2019gia,Wang:2019skw,Tian:2019yhn,Banerjee:2019nnj,Shaikh:2019hbm,Kumar:2019pjp,Allahyari:2019jqz,Li:2019lsm,Jusufi:2019ltj,Rummel:2019ads,Kumar:2020hgm,Vagnozzi:2020quf,Li:2020drn,Narang:2020bgo,Liu:2020ola,Konoplya:2020bxa,Guo:2020zmf,Pantig:2020uhp,Wei:2020ght,Kumar:2020owy,Roy:2020dyy,Islam:2020xmy,Chen:2020aix,Jin:2020emq,Sau:2020xau,Jusufi:2020dhz,Creci:2020mfg,Kumar:2020oqp,Chen:2020qyp,Zeng:2020dco,Neves:2020doc,Badia:2020pnh,Jusufi:2020cpn}.


Having established in the above paragraphs that exploring controlled violations of the NHT, particularly in the context of charged hairy BHs, is timely and interesting beyond the field of gravitational physics, in this work we shall also venture alontg this route. One approach towards studying violations of the NHT is to build parametrized families of metrics deviating from the GR BH metrics (Schwarzschild, Kerr, Reissner-Nordstr\"{o}m, or Kerr-Newman), which are flexible but not necessarily tied to any specific theory beyond GR, whereas another approach consists of searching for exact solutions of a coupled gravity-matter system yielding deviations from the aforementioned metrics. It is the latter approach we shall follow in this work. In particular, we shall focus on two specific models where the gravitational sector is described by GR, and the matter field features a scalar field in the presence of an electromagnetic field. Therefore, we shall consider a system described by the Einstein-Hilbert-Maxwell-Klein-Gordon equations. Earlier works on charged BHs with scalar hair can be found for instance in~\cite{Martinez:2006an,Anabalon:2009qt,Xu:2013nia,Anabalon:2013qua,Fan:2015oca}.

The first model we study features a minimally-coupled scalar field with a potential in the presence of an electromagnetic field. The second model instead features a scalar field conformally-coupled to gravity, also in the presence of an electromagnetic field. For both models we shall construct charged BH solutions with scalar hair, which thus go beyond the charged non-rotating Reissner-Nordstr\"{o}m BH solution of GR. We will refer to the resulting BHs as minimally-coupled charged hairy BH (MCCHBH) and conformally-coupled charged hairy BH (CCCHBH) respectively. We will then compute the shadows of these two BHs, and then confront such shadows with the shadow of M87* imaged by the EHT collaboration. In doing so, we will extract constraints on the fundamental parameters of our two models. In this work we shall ignore the BH spin, \textit{i.e.} the BH solutions we will construct are non-rotating. We will argue that introducing rotation does not alter our results in a significant manner, and will leave the construction of rotating charged hairy BHs to future projects.

The rest of this paper is then organized as follows. In Section~\ref{sec:mcchbh} we consider the first model (a minimally-coupled scalar field with a potential in the presence of an electromagnetic field) and construct the relevant BH solution (which we refer to as MCCHBH). In Section~\ref{sec:shadowmcchbh} we then compute the shadow of the MCCHBH. In Section~\ref{sec:ccchbh} we instead consider the second model (a conformally-coupled scalar field in the presence of an electromagnetic field), and construct the relevant BH solution (which we refer to as CCCHBH). In Section~\ref{sec:shadowccchbh} we compute the shadow of the CCCHBH. In Section~\ref{sec:ehtshadow} we compare the shadows of the two classes of black holes we have studied against the shadow of M87* imaged by the Event Horizon Telescope, and discuss some important caveats of our study. Finally, we provide concluding remarks in Sec.~\ref{sec:conclusions}. Throughout this paper we work with Planck units wherein $c=\hbar=G=1$.

\section{Minimally-coupled charged black holes with scalar hair: MCCHBH}
\label{sec:mcchbh}

We begin by considering an action whose gravitational sector consists of General Relativity (GR), supplemented by a minimally-coupled scalar field $\phi$ with a potential $V(\phi)$, in the presence of an electromagnetic field with field-strength tensor given by $F_{\mu \nu}$. The Einstein-Hilbert-Maxwell-Klein-Gordon action we envisage is thus given by:
\begin{eqnarray}
S=\int d^{4}x\sqrt{-g}\left(\frac{1}{2\kappa}R-\frac{1}{4}F_{\mu \nu}F^{\mu \nu} -\frac{1}{2}g^{\mu\nu}\nabla_{\mu}\phi\nabla_{\nu}\phi-V(\phi)\right)\,,
\label{eq:smcchbh}
\end{eqnarray}
where $\kappa \equiv 8\pi$, the electromagnetic field has 4-potential given by $A_{\mu}=(A_t(r),0,0,0)$ and field-strength tensor given by $F_{\mu\nu}=\partial_\mu\,A_\nu-\partial_{\nu}A_\mu$. Related models have been considered earlier in e.g.~\cite{Martinez:2006an,Anabalon:2009qt,Xu:2013nia,Anabalon:2013qua,Fan:2015oca,Gonzalez:2013aca,Gonzalez:2014tga,Hui:2019aqm}.

Let us now consider the field equations which follow from the action given in Eq.~(\ref{eq:smcchbh}). The gravitational field equations are obtained by extremizing the action with respect to the metric tensor, and are given by:
\begin{eqnarray}
R_{\mu\nu}-\frac{1}{2}g_{\mu\nu}R=\kappa (T^{(\phi)}_{\mu\nu}+T^{(F)}_{\mu\nu})\,,
\label{eq:einsteinhilbert}
\end{eqnarray}
where the matter (right-hand side) section of the Einstein-Hilbert equations is sourced by the scalar and electromagnetic fields, whose energy-momentum tensors ($T^{(\phi)}_{\mu\nu}$ and $T^{(F)}_{\mu\nu}$ respectively) are given by the following:
\begin{eqnarray}
T^{(\phi)}_{\mu\nu} &=&\nabla_{\mu}\phi\nabla_{\nu}\phi-g_{\mu\nu} \left [ \frac{1}{2}g^{\rho\sigma}\nabla_{\rho}\phi\nabla_{\sigma}\phi+V(\phi) \right ] \,,\\
T^{(F)}_{\mu\nu} &=& F_{\mu}^{\alpha}F_{\nu \alpha} -\frac{1}{4}g_{\mu \nu}F_{\alpha\beta}F^{\alpha\beta}\,,
\label{eq:energymomentum}
\end{eqnarray}
where of course since $\phi$ is a scalar field, $\nabla_{\mu}\phi = \partial_{\mu}\phi$. Extremizing the action with respect to the scalar field gives us the usual Klein-Gordon equation:
\begin{eqnarray}
\Box\phi =\frac{dV}{d\phi}\,.
\label{eq:kleingordon}
\end{eqnarray}
Finally, extremizing the action with respect to the 4-potential gives us the usual Maxwell equations:
\begin{eqnarray}
\nabla_{\nu}F^{\mu\nu}=0\,.
\label{eq:maxwell}
\end{eqnarray}
Therefore, Eqs.~(\ref{eq:einsteinhilbert}-\ref{eq:maxwell}) fully specify the dynamics of the system.

We shall now seek static spherically symmetrically (SSS) solutions to the field equations. In general, astrophysical systems will rotate, and thus it is usually not possible to neglect angular momentum. In doing so, the solutions we will find should be seen as approximate descriptions of realistic astrophysical systems. However, as we shall argue later in Sec.~\ref{sec:ehtshadow}, this approximation can be justified given the current level of observational precision in VLBI shadow imaging. Our ansatz for the SSS squared line element is given by:
\begin{eqnarray}
ds^{2}=-f(r)dt^{2}+f^{-1}(r)dr^{2}+a^{2}(r)d\sigma^2\,,
\label{metricBH}
\end{eqnarray}
where $d \sigma ^2$ is the metric of the spatial 2-section, with positive curvature. Plugging in our squared line element from Eq.~(\ref{metricBH}) into Eqs.(\ref{eq:einsteinhilbert},\ref{eq:energymomentum}), some tedious but straightforward algebra leads us to the following three independent differential equations:
\begin{eqnarray}
\label{first}
f''(r)+2\frac{a'(r)}{a(r)}f'(r)+2V(\phi)&=&A^{\prime}_t(r)^2\,,\\
\label{second}
\frac{a'(r)}{a(r)}f'(r)+ \left ( \left ( \frac{a'(r)}{a(r)} \right )^{2}+\frac{a''(r)}{a(r)} \right ) f(r)-\frac{1}{a(r)^{2}}+V(\phi)&=&-\frac{1}{2}A^{\prime}_t(r)^2\,,\\
\label{third}
f''(r)+2\frac{a'(r)}{a(r)}f'(r)+ \left ( 4\frac{a''(r)}{a(r)}+2(\phi'(r))^{2} \right ) +2V(\phi)&=&A^{\prime}_t(r)^2\,.
\end{eqnarray}
To make progress, we replace $ V(\phi)$ from Eq.~(\ref{first}) into Eqs.~(\ref{second},\ref{third}), from which we get:
\begin{eqnarray}
\label{adiff}
a''(r)+\frac{1}{2}(\phi'(r))^2 a(r)&=&0\,,\\
\label{fdiff}
f''(r)-2 \left ( \left ( \frac{a'(r)}{a(r)} \right ) ^{2}+\frac{a''(r)}{a(r)} \right ) f(r)+\frac{2}{a(r)^2}&=&2A^{\prime}_t(r)^2\,,
\end{eqnarray}
The field Klein-Gordon and Maxwell equations, Eqs.~(\ref{eq:kleingordon},\ref{eq:maxwell}), in addition to the differential equations Eqs.~(\ref{first}-\ref{third}), are the ingredients we need to extract the hairy BH solutions (recall we refer to the hairy BH solutions within this theory as MCCHBH).

We now have to make a well-behaved ansatz for the scalar field, which should be regular at the horizon and fall off sufficiently quickly at infinity. Following~\cite{Gonzalez:2013aca,Gonzalez:2014tga} we will work under the following ansatz:
\begin{eqnarray}
\phi(r) =\frac{1}{\sqrt{2}}\ln \left ( 1+\frac{\nu}{r} \right )\,,
\label{field}
\end{eqnarray}
where we have introduced a scalar charge parameter $\nu$ with carries dimensions of length. Recall we discussed in Sec.~\ref{sec:intro} that one way to evade the NHT in a controlled way is by introducing a new scale, via a cosmological constant and/or in the scalar sector. Explicitly introducing the scalar charge parameter $\nu$ follows the latter approach. The ansatz we made in Eq.~(\ref{field}) is well-motivated, see e.g. discussions in~\cite{Gonzalez:2013aca,Gonzalez:2014tga} and references to these works. Plugging the ansatz Eq.~(\ref{field}) into Eqs.~(\ref{eq:kleingordon},\ref{adiff},\ref{fdiff}), we can analytically solve for the 0-component of the 4-potential, $A_t(r)$, as well as the function $a(r)$ appearing in the line element Eq.~(\ref{metricBH}). We find that these two quantities are given by:
\begin{eqnarray}
\nonumber a\left( r\right)& =&\sqrt{r\left( r+\nu \right) }\,,\\
A_t(r)&=&\frac{Q}{\nu}\ln\left(\frac{r}{r+\nu}\right)\,,
\label{ametric}
\end{eqnarray}
where $Q$ is the electric charge of the BH as resulting from Maxwell's equations, note that $\lim_{\nu\rightarrow0}A_t(r)=-Q/r$. Similarly, we find that the metric function $f(r)$ is given by:
\begin{eqnarray}
f(r) &=&-2\frac{Q^2}{\nu^2}+C_1r(r+\nu)-\frac{C_2(2r+\nu)}{\nu^2}+2\frac{r(2r+\nu)}{\nu^2} \nonumber \\
&& -2 \left ( \frac{Q^2(2r+\nu)+r(r+\nu)(C_2+\nu)}{\nu^3}+\frac{Q^2r(r+\nu)\ln\frac{r}{r+\nu}}{\nu^4} \right ) \ln\frac{r}{r+\nu}\,,
\label{f(r)}
\end{eqnarray}
where $C_1$ and $C_2$ are integration constants.

Note that so far we had not specified anything about the scalar field potential. In fact, the usual approach is to first define the potential, and from that solve for the scalar field profile. Instead, we have started by making an ansatz for the scalar field profile, motivated by the necessity of having a well-behaved (regular and falling fast enough at infinity) scalar field profile. Therefore, we now have to work backwards to understand what is the potential which can give rise to the given scalar field profile.~\footnote{This approach is similar to the reconstruction procedure often applied in cosmology to reconstruct the scalar field potential which can give rise to a chosen expansion history and/or observational indices, see for instance~\cite{Rinaldi:2014gua,Myrzakulov:2015fra,Odintsov:2018zhw}.} We can do so by inserting Eq.~(\ref{f(r)}) into Eq.~(\ref{first}), from which we get that the potential is given by the following:
\begin{eqnarray}
\nonumber V(\phi)=\frac{1}{2\nu^4}e^{-2\sqrt{2}\phi} g(\phi)\,,
\label{m2potn}
\end{eqnarray}
where the function $g(\phi)$ is given by:
\begin{eqnarray}
g(\phi) &=& \left ( e^{\sqrt{2}\phi}-1\right)^2 \left ( 1+10e^{\sqrt{2}\phi} +e^{2\sqrt{2}\phi} \right ) Q^2 \nonumber \\
&& +e^{\sqrt{2}\phi}\nu \left ( -6C_2-10\nu-C_1\nu^3-4e^{\sqrt{2}\phi}\nu \left ( 4+C_1\nu^2 \right ) +e^{2\sqrt{2}\phi} \left ( 6C_2+2\nu-C_1\nu^3 \right ) \right ) \nonumber	\\
&& +2\sqrt{2}\phi \left ( \left ( 1+4e^{\sqrt{2}\phi}+e^{2\sqrt{2}\phi} \right ) Q^2\ln\frac{\nu}{e^{\sqrt{2}\phi-1}} \right. \nonumber \\
&& +2e^{\sqrt{2}\phi} \left. \left ( \left ( 2+\cosh\sqrt{2}\phi \right ) \left ( \nu C_2+\nu^2-Q^2\ln\frac{e^{\sqrt{2}\phi}\nu}{e^{\sqrt{2}\phi-1}} \right ) +6Q^2\sinh\sqrt{2}\phi \right ) \right )
\label{eq:gphi}
\end{eqnarray}
In the following, rather than considering asymptotic dS or AdS space-time, we will be interested in asymptotically flat space-time, \textit{i.e.} with $\Lambda=0$. Therefore, we impose appropriate boundary conditions such that $V(0) =\Lambda_{\rm eff} \to 0$. Following~\cite{Gonzalez:2013aca,Gonzalez:2014tga}, we fix $C_1=-4/\nu^2$.

The final step consists of re-inserting this value of $C_1$ into the metric function $f(r)$, and additionally imposing that in the limit $\nu \to 0$, the metric function Eq.~(\ref{f(r)}) reduces to that of the standard Reissner-Nordstr\"{o}m (RN) metric for a charged non-rotating BH. If we are able to satisfy this condition, we would ensure that the BH solution we have found carries primary hair. Doing so we find that the metric function, which we refer to as $f_{\rm MC}(r)$ (where ``MC'' stands for ``minimally-coupled'', and recall that we refer to the resulting minimally-coupled charged BH with scalar hair as MCCHBH), is given by:
\begin{eqnarray}
f_{\rm MC}(r) &=&-\frac{2Q^2+2r\nu +6M(2r+\nu)}{\nu^2} \nonumber\\
&& - \left ( \frac{ 2Q^2(2r+\nu)+2r(r+\nu)(6M+\nu)}{\nu^3}+\frac{2Q^2r(r+\nu)\ln\frac{r}{r+\nu}}{\nu^4} \right ) \ln\frac{r}{r+\nu}\,.
\label{f(rr)}
\end{eqnarray}
From now on, we shall fix $M=1$, and thus work in units of the MCCHBH mass. Note that in moving from Eq.~(\ref{f(r)}) to Eq.~(\ref{f(rr)}) we have set $C_2=6M$, by requiring that in the hairless limit $\nu \rightarrow 0$ the metric function reduces to that of the standard RN metric.

Before moving forward, we can perform a few sanity checks. First of all, we can make sure that the charged metric function for the MCCHBH in Eq.~(\ref{f(rr)}) is consistent by replacing the above metric function into Eq.~(\ref{fdiff}). We have explicitly verified that the metric function we have found satisfies Eq.~(\ref{fdiff}), with the 0-component of the 4-potential given by Eq.~(\ref{ametric}). Moreover, it is easy to see that in the limit $\nu \to 0$ we recover the standard RN metric, confirming that our MCCHBH carries primary hair. Furthermore, it is easy to check that the metric function in Eq.~(\ref{f(rr)}) is well-behaved at large distances, since $\lim_{r \to \infty}f_{\rm MC}(r)=1$, as one would expect given that in this regime the scalar field decouples. However, unlike the standard RN BH, the metric function Eq.~(\ref{f(rr)}) has no extremal limit since whenever $\nu \neq 0$ the condition $f_{\rm MC}(r)=f_{\rm MC}'(r)=0$ is no longer satisfied. Therefore, we end up with separate Cauchy and event horizons, which we denote by $r_c$ and $r_e$ respectively. Moreover, values $Q>1$ are now allowed, unlike for RN BHs, for which $Q=1$ characterizes extremal BHs.

\section{Shadows of minimally-coupled charged black holes with scalar hair}
\label{sec:shadowmcchbh}

Having determined the metric of the minimally-coupled charged black hole with scalar hair (MCCHBH), we can now proceed to compute the shadow it casts. Let us begin by considering photon geodesics parametrized by $x^{\mu}(\tau)$ in terms of an affine parameter $\tau$. On the space-time metric given by Eq.~(\ref{metricBH}), the Lagrangian of the photon geodesics can be written as:
\begin{eqnarray}
\mathcal{L}=-f_{\rm MC}(r)\dot t^2+\frac{1}{f_{\rm MC}(r)}\dot r^2 + a^2 \dot \theta^2+a^2 \dot\phi^2\,,
\label{eff1}
\end{eqnarray}
where the dot denotes differentiation with respect to $\tau$. Since we are considering spherically symmetric solutions, we can safely restrict our attention to orbits confined to the equatorial plane (\textit{i.e.} $\theta=\pi/2$). The equations of motion for a null geodesic confined to the equatorial plane are given by the following:
\begin{eqnarray}
\label{eq:e}
&&E=f_{\rm MC}(r) \,\dot{t}\,,\\
\label{eq:l}
&&L=a^2\dot{\phi}\,,\\
\label{R}
&&f_{\rm MC}(r)\dot{t}^2-\frac{1}{f_{\rm MC}(r)} \left(\frac{dr}{d\phi} \right)^2\dot\phi^2-a^2 \dot\phi^2=0\,.
\end{eqnarray}
where $E$ and $L$ are the photon's total energy and angular momentum, both of which are constants of motion along the geodesic. Plugging Eqs.~(\ref{eq:e},\ref{eq:l}) into Eq.~(\ref{R}) we can express the equation for the radial coordinate as:
\begin{eqnarray}
\left(\frac{dr}{d\phi} \right)^2=U(r)=a^4(r) \left(-\frac{f_{\rm MC}(r)}{a^2}+\frac{E^2}{L^2} \right)\,,
\label{V}
\end{eqnarray}
where $U(r)$ can be interpreted as an effective potential.

To study the photon sphere of MCCHBHs, which will characterize the shadows they cast, we need to focus on unstable circular orbits. We find these orbits by demanding:
\begin{eqnarray}
U(r) = \frac{dU(r)}{dr} = 0\,, \quad \frac{d^2U(r)}{dr^2}>0\,.
\label{eq:unstablecircularorbits}
\end{eqnarray}
Using the potential given in Eq.~(\ref{V}), the conditions for unstable circular orbits given in the first of Eq.~(\ref{eq:unstablecircularorbits}) translate to:
\begin{eqnarray}
\label{b}
&&b^2=\frac{L^2}{E^2}=\frac{a^2}{f_{\rm MC}(r)}\,,\\
\label{un}
&&2\frac{da}{dr}f_{\rm MC}(r)-a(r)\frac{df_{\rm MC}(r)}{dr}=0\,.
\end{eqnarray}
The quantity $b$ is referred to as \textit{impact parameter}, and is directly related to the size of the BH shadow. We can insert the metric function for the MCCHBH, Eq.~(\ref{f(rr)}), into Eqs.~(\ref{b},\ref{un}) and (\ref{un}), from which we get the following two equations:
\begin{eqnarray}
\label{b1}
&b= \left (-\frac{2\nu ^2 (3 \nu +6 r+2 Q^2+\nu  r)+2\nu \ln \left ( \frac{r}{\nu +r} \right ) \left ( r (\nu +6) (\nu +r)+2 Q^2 (\nu +2 r) \right ) +4 Q^2 r (\nu +r) \ln ^2 \left ( \frac{r}{\nu+r} \right ) }{\nu ^4 r(r+\nu)} \right ) ^{-1/2}\,, \\
\label{un1}
&\frac{2 \nu  (r-3M)-2 Q^2 \ln \left(\frac{r}{\nu +r}\right)}{\nu\sqrt{r(r+\nu)}}=0\,,
\end{eqnarray}
As a sanity check, we can set $Q=0$ and $\nu=0$. In this case we find that Eqs.~(\ref{b1},\ref{un1}) describe an unstable critical curve located at $r_{\rm ph-sch} = 3$, with the relevant critical impact parameter being given by $b_{\rm ph-sch} = 3\sqrt{3}$ (recall that we are working in units of MCCHBH mass, \textit{i.e.} we set $M=1$). Both results agree with the standard results for an uncharged hairless Schwarzschild BH.

To make progress, we have to solve the above equations. We solve Eq.~(\ref{un1}) numerically. For purely pedagogical purposes, in Tab.~\ref{Nu} we show our numerical solutions of Eqs.~(\ref{f(rr)},\ref{un1}) for two representative values of the hair parameter, $\nu=0\,,2$. In particular, for each of the two values of the hair parameter (the first of the two of course corresponding to a hairless BH), we consider various values of the BH charge $Q$, and for each of them we compute the radial coordinate of the Cauchy horizon $r_c$, the radial coordinate of the event horizon $r_e$, and the radial coordinate of the photon sphere $r_{ph}$. Recall that all three the radial coordinates, as well as the hair parameter $\nu$ which has dimensions of length, are measured in units of the MCCHBH mass $M$. As expected, for the hairless case only solutions with $Q<1$ are allowed. $Q=1$ constitutes the extremal RN case, for which the Cauchy and event horizon coincide, and the photon sphere is located at $r_{ph}=2M$. If the hair parameter is non-zero, as we anticipated earlier, values $Q>1$ are allowed.

\begin{table}
\begin{center}
\begin{tabular}{|c|c|c|c||c|c|c|c|}
\hline
$Q$&$r_c(\nu=0)$&$r_{e}(\nu=0)$&$r_{ph}(\nu=0)$&$r_c(\nu=2)$&$r_{e}(\nu=2)$&$r_{ph}(\nu=2)$\\
\hline
$0$&$-$&$2$ &$3$&$-$&$1.742$ &$3$ \\  \hline
$0.1$&$0.005$&$1.995$ &$2.99332$&$\ll 10^{-3}$&$1.740$ &$2.997$\\  \hline
$0.2$&$0.020$&$1.980$ &$2.97309$&$\ll 10^{-3}$&$1.735$ &$2.990$\\ \hline
$0.3$&$0.046$&$1.954$ &$2.93875$&$\ll 10^{-3}$&$1.725$ &$2.977$\\ \hline
$0.4$ &$0.083$& $1.917$ &$2.88924$&$\ll 10^{-3}$&$1.712$ &$2.959$ \\ \hline
$0.5$&$0.133$&$1.866$ &$2.82287$&$\ll 10^{-3}$&$1.695$ &$2.935$\\ \hline
$0.6$&$0.200$&$1.800$ &$2.73693$&$\ll 10^{-3}$&$1.673$ &$2.906$ \\ \hline
$0.7$&$0.286$&$1.714$ &$2.62694$&$\ll 10^{-3}$&$1.647$ &$2.870$ \\ \hline
$0.8$&$0.400$&$1.600$ &$2.48489$&$\ll 10^{-3}$&$1.617$ &$2.829$\\ \hline
$0.9$&$0.564$&$1.436$ &$2.29373$&$\ll 10^{-3}$&$1.581$ &$2.781$\\ \hline
$1$&$1$&$1$ &$2$&$0.002$&$1.540$ &$2.725$\\ \hline
$1.1$&$-$&$-$ &$-$&$0.006$ &$1.492$ &$2.661$ \\ \hline
$1.2$&$-$&$-$ &$-$&$0.014$ &$1.437$ &$2.588$\\ \hline
$1.3$&$-$&$-$ &$-$&$0.029$ &$1.373$ &$2.504$\\ \hline
$1.4$&$-$&$-$ &$-$&$0.055$ &$1.299$ &$2.407$\\ \hline
$1.5$&$-$&$-$ &$-$&$0.095$ &$1.210$ &$2.295$\\ \hline
$1.6$&$-$&$-$ &$-$&$0.158$ &$1.101$ &$2.162$\\ \hline
$1.7$&$-$&$-$ &$-$&$0.261$ &$0.954$ &$1.998$\\ \hline
$1.8$&$-$&$-$ &$-$&$0.524$ &$0.649$ &$1.780$\\ \hline
\end{tabular}
\caption{Numerical solutions of Eqs.~(\ref{f(rr)},\ref{un1}) for two representative values of the hair parameter of the minimally-coupled charged black hole with scalar hair (MCCHBH), $\nu=0\,,2$. For each of the two values of the hair parameter, we consider various values of the BH charge $Q$, and report $r_c$, $r_e$, and $r_{ph}$, the radial coordinates of the Cauchy horizon, event horizon, and photon sphere respectively. Recall that all three radial coordinates, the hair parameter $\nu$ which has dimensions of length, and the charge $Q$ are measured in units of the MCCHBH mass $M$. Note that the hightest value $Q=1.8$ we report does not result in an extremal charged BH, since $r_c \neq r_e$.}
\label{Nu}
\end{center}
\end{table}

Our results show, quite unexpectedly, that by turning on the scalar hair parameter $\nu$, the radial coordinates of all three the Cauchy horizon $r_c$, the event horizon $r_e$, and the photon sphere $r_{ph}$ are modified with respect to the hairless case. Turning on the scalar hair parameter always makes $r_c$ decrease. For sufficiently small electric charge values, $Q \lesssim 0.7$, we find that $r_e$ decreases as well, although it increases for larger $Q$. Finally, $r_{ph}$ appears to increase regardless of the value of $Q$.

Finally, let us consider the four-vector $K^{\mu}$ tangent to the photon's path, which using Eqs.~(\ref{eq:e},\ref{eq:l},\ref{V}) is given by::
\begin{eqnarray}
K^{\mu}=\frac{dx^{\mu}}{d\tau}=\left(\frac{a^2}{b f_{\rm MC}(r)} ,\sqrt{U(r)},0,1\right)\,.
\label{eq:tangent}
\end{eqnarray}
We want to compute the angle $\psi$ between $K^{\mu}$ and $D^{\mu}$, with $D^{\mu}=\left(0,r,0,0 \right)$ the 4-vector of a static distant observer located at $r=r_0$. The angle $\psi$ is given by~\cite{Rindler:2007zz}:
\begin{eqnarray}
\psi=\cos^{-1} \left ( \sqrt{\dfrac{ U(r)}{U(r)+f_{\rm MC}(r) a^2}} \right ) \,.
\label{psi}
\end{eqnarray}
A few comments on how the value of $\psi$ in Eq.~(\ref{psi}) is used to compute the shadow size are in order. In particular, there are two main approaches towards computing the sizes of BH shadows in the literature. In the first approach, one projects the photon four-vector to the observer's frame using suitable tetrads at infinity. We use a second approach, similar to that used in~\cite{Firouzjaee:2019aij,Chang:2020miq}. In asymptotically flat space-times, for observers located at infinity or sufficiently far, the two methods return the same answer. In fact, in the absence of rotation, we expecte that the shadow of the BH is a perfect circle with coordinates obeying the equation $X^2+Y^2=R_S^2$, where $(X, Y)=(R_S \cos w, R_S \sin w )$ with $w \in [0,2\pi]$. Therefore, $R_S$ is understood to be the shadow's radius and is given by $r_0 \tan \psi$ for a distant observer. Note that we are considering asymptotically flat space-time. See e.g.~\cite{Perlick:2018iye,Bisnovatyi-Kogan:2018vxl,Tsupko:2019pzg} for important works on shadows of BHs in asymptotically de Sitter space-times.

\begin{figure}[!ht]
\begin{center}
\includegraphics[width=0.45\linewidth]{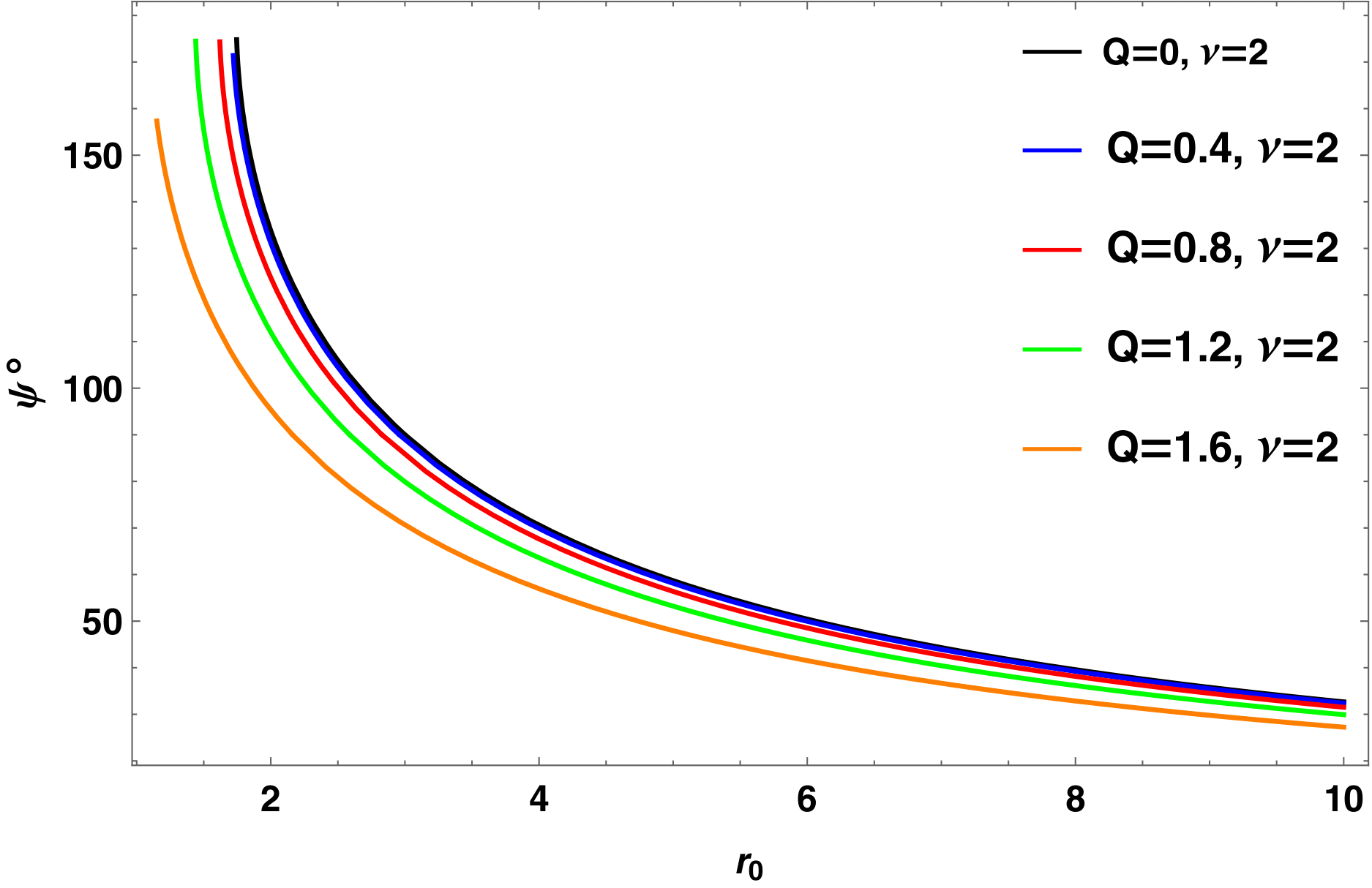}
\includegraphics[width=0.45\linewidth]{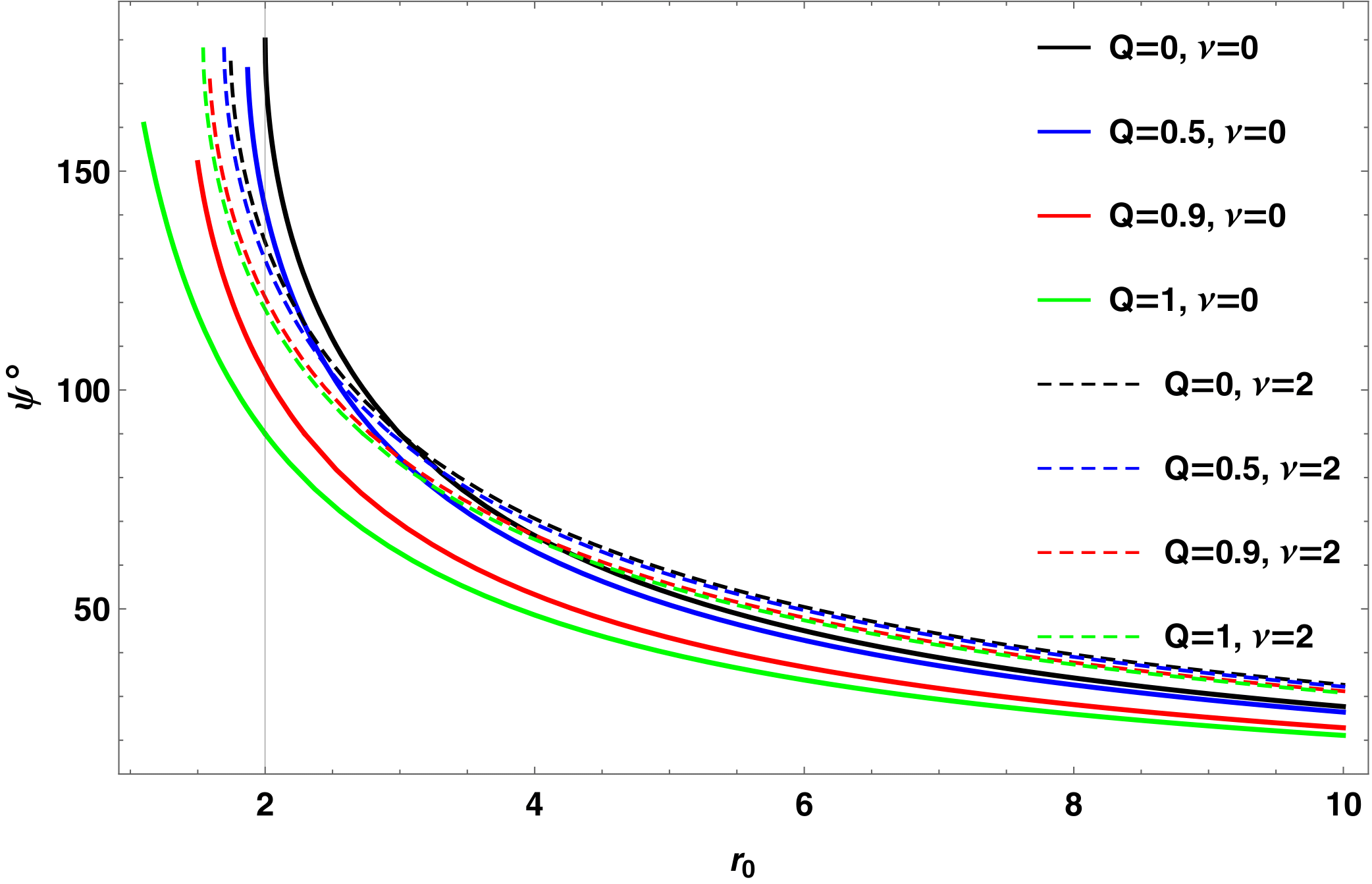} \\
\includegraphics[width=0.45\linewidth]{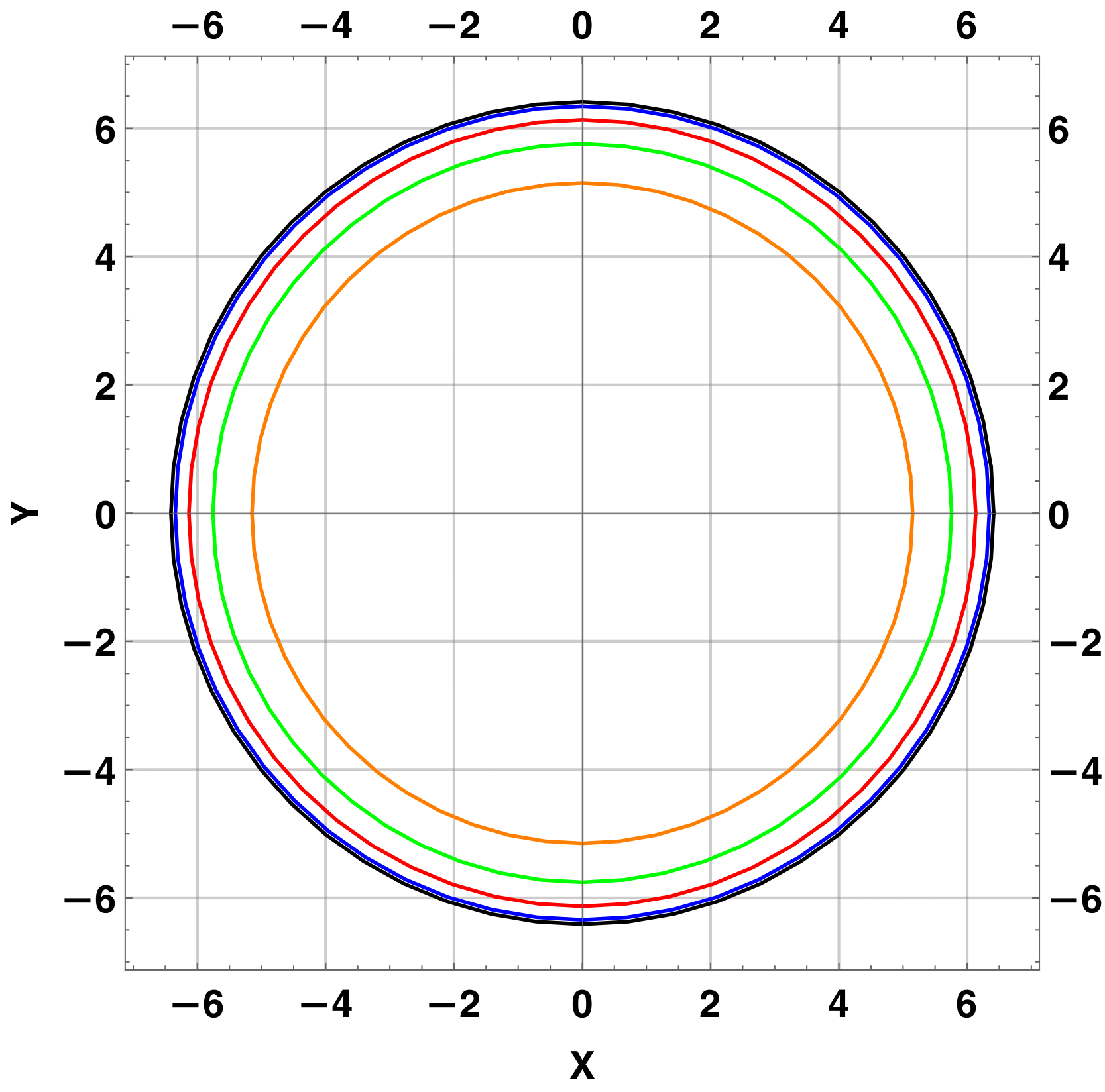}
\includegraphics[width=0.45\linewidth]{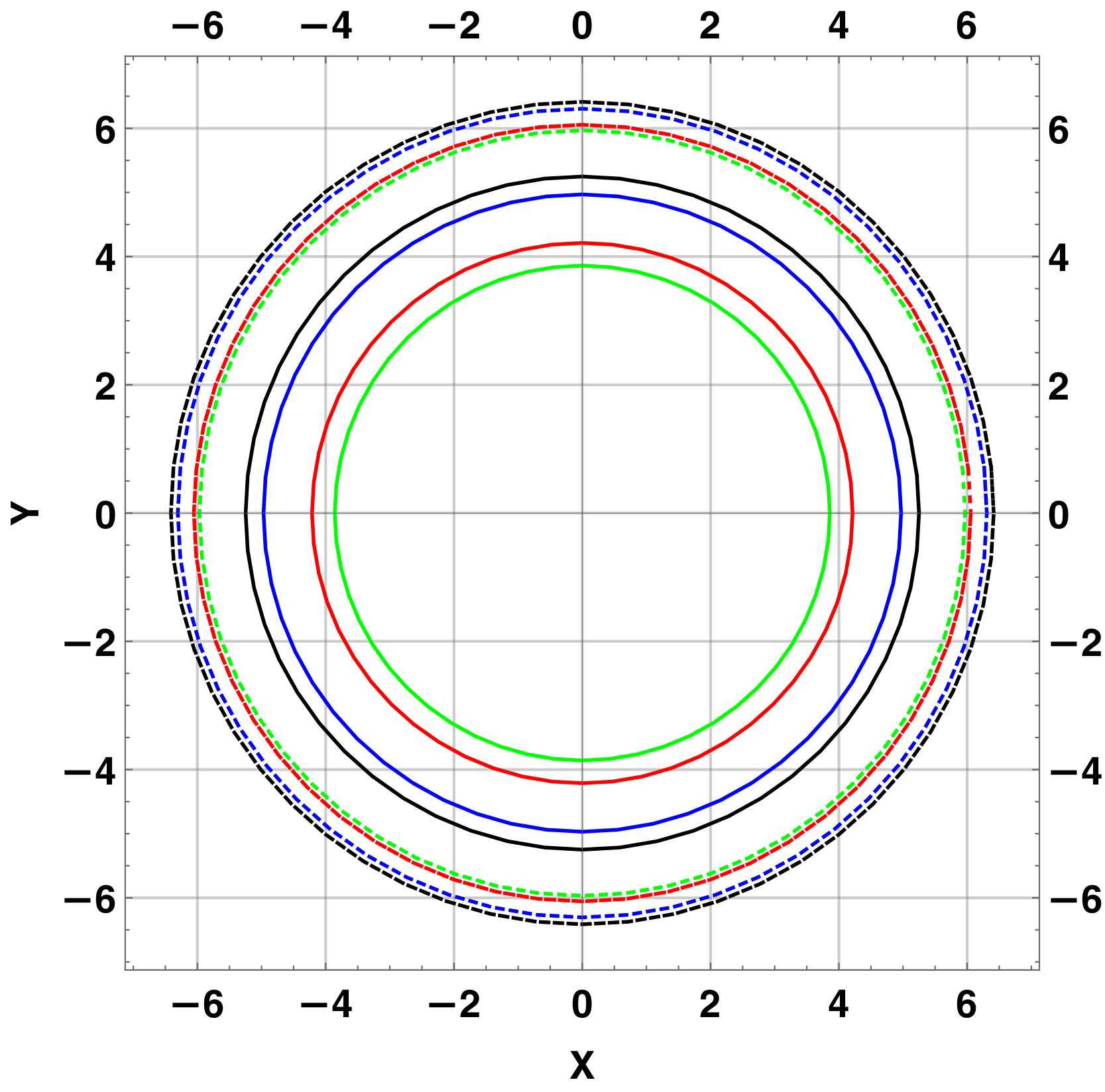}
\caption{\textit{Upper panels}: plot of the angle between the four-vector tangent to the path of a photon on the photon sphere and the position coordinate of a static distant observer, $\psi$, versus the radial coordinate of a distant observer $r$, for several values of the electric charge $Q$ and scalar hair parameter $\nu$ of the minimally-coupled charged hairy black holes (MCCHBHs) studied in this section. In particular, for any given value of $Q$ (represented by a given color), the right panel allows the comparison between a MCCHBH (dashed curves) and a Reissner-Nordstr\"{o}m BH (solid curves) with the same electric charge $Q$. \textit{Lower panels}: shadows cast by the MCCHBH, as seen by a static distant observer. The underlying values of the electric charges $Q$ and scalar hair parameter $\nu$ follow the same colour-coding as the upper panels, with the right panel allowing for a comparison between a MCCHBH (dashed curves) and a Reissner-Nordstr\"{o}m BH (solid curves) with the same electric charge $Q$.}
\label{SH1}
\end{center}
\end{figure}

By making use of our numerical solution, of which we gave some representative examples in Tab.~\ref{Nu}, we can now compute the MCCHBH shadow as a function of the hair parameter $\nu$. In Fig.~\ref{SH1} we plot both the angle $\psi$ as a function of $r_0$ for various choices of the electric charge $Q$ and hair parameter $\nu$ (upper panels), and the resulting MCCHBH shadows (lower panels).

The lower panels of Fig.~\ref{SH1} illustrate a few important points. As we see from the lower left panel, where we plot the shadow at a fixed value of the hair parameter $\nu$ but varying the value of the electric charge $Q$, increasing the electric charge at fixed hair parameter $\nu=2$ makes the shadow shrink. This is exactly the same behaviour one sees for the standard Reissner-Nordstr\"{o}m BH ($\nu=0$), where increasing the electric charge also makes the shadow shrink. In the lower right panel, we can compare the hairless Reissner-Nordstr\"{o}m BH (solid curves) against a hairy BH (dashed curves) with the same electric charge. We see that the effect of the hair parameter is to make the shadow larger. The effect is that, even increasing $Q$ (which would on its own shrink the shadow, as we discussed above), the size of the hairy BH shadow can in principle be larger than that of the Schwarzschild BH for which $r=3\sqrt{3}$. Therefore, we can expect that observations of BH shadows at known BH mass can in principle set an upper limit on the hair parameter. The results shown in Fig.~\ref{SH1} confirm the expectations from our numerical results shown in Tab.~\ref{Nu}.

\section{Conformally-coupled charged black holes with scalar hair and their shadows: CCCHBH}
\label{sec:ccchbh}

In the previous part of the work we focused on BHs within the context of GR supplemented by a minimally-coupled scalar field with a potential in the presence of an electromagnetic field. This time we will instead consider a different theory, still featuring an additional scalar field $\phi$. The action we consider is the following:
\begin{eqnarray}S_{\rm}=\frac{1}{2\kappa}\int d^4 x\sqrt{-g} \left [ R-F_{\mu\nu}F^{\mu\nu}-\dfrac{\kappa}{6} \left ( \phi^2R+ 6\partial_\mu\phi\partial^\mu\phi \right ) \right ] \,,
\label{EMCS}
\end{eqnarray}
where we have conformally coupled the scalar field to gravity. Black hole or black brane solutions in similar models have been studied earlier in e.g.~\cite{Martinez:2005di,Cisterna:2018hzf,Myung:2018jvi,Myung:2019oua,Myung:2019adj,
Zou:2019ays,Zou:2020rlv,Caceres:2020myr}. The gravitational field equations associated to the action in Eq.~(\ref{EMCS}) are given by:
\begin{eqnarray}
R_{\mu\nu}-\frac{R}{2}g_{\mu\nu}=2T^{\rm M}_{\mu\nu}+T^{\rm \phi}_{\mu\nu}\,,
\label{nequa1}
\end{eqnarray}
where the energy-momentum tensors for the scalar and electromagnetic fields, $T^{(\phi)}_{\mu\nu}$ and $T^{(M)}_{\mu\nu}$ respectively, are given by the following:
\begin{eqnarray}
\label{equa2}
T^{(\phi)}_{\mu\nu}&=&\dfrac{\kappa}{6} \left ( \phi^2(R_{\mu\nu}-\frac{R}{2}g_{\mu\nu})+g_{\mu\nu}\nabla^2(\phi^2)- \nabla_\mu\nabla_\nu(\phi^2)+6\nabla_\mu\phi\nabla_\nu\phi-3(\nabla\phi)^2g_{\mu\nu} \right ) \,,\\
\label{trace}
T^{({\rm M})}_{\mu\nu}&=&F_{\mu\rho}F_{\nu}^\rho-\frac{F^2}{4}g_{\mu\nu}\,.\\
\end{eqnarray}
respectively. The evolution of the electromagnetic field is given as usual by Maxwell's equations:
\begin{eqnarray}
\nabla^\mu F_{\mu\nu}=0\,.
\label{maxwell-eq}
\end{eqnarray}
The scalar field equations take the form:
\begin{eqnarray}
\left ( \Box-\frac{R}{6} \right ) \phi=0\,,
\label{ascalar-eq}
\end{eqnarray}
However, as noted earlier in~\cite{Zou:2019ays,Zou:2020rlv}, things simplify considerably on-shell. While $T^{\rm (M)}_{\mu\nu}$ is traceless (\textit{i.e.} $T^{({\rm M})\mu}_{\mu}=0$), the trace of $T^{(\phi)}_{\mu\nu}$ reads:
\begin{eqnarray}
T^{(\phi)\mu}_{\mu} = -\phi \left ( R\phi - 6\Box\phi \right )\,,
\label{eq:trace}
\end{eqnarray}
which manifestly vanishes on-shell when Eq.~(\ref{ascalar-eq}) holds. Then, taking the trace of Eq.~(\ref{nequa1}) on-shell gives $R=0$. Plugging this back into Eq.~(\ref{ascalar-eq}) simplies the equation to:
\begin{eqnarray}
\Box \phi = 0
\label{eq:conformal}
\end{eqnarray}
In conclusion, Eqs.~(\ref{nequa1},\ref{maxwell-eq},\ref{eq:conformal}) fully specify the dynamics of the system.

As in Sec.~\ref{sec:mcchbh}, we shall now seek static spherically symmetric solutions of the equations of motion. We consider once more a line element given by:
\begin{eqnarray}
ds^{2}=-f(r)dt^{2}+f^{-1}(r)dr^{2}+r^{2}d \sigma^2\,.
\label{metric}
\end{eqnarray}
The full solution (metric function, which we shall denote by $f_{\rm CC}(r)$, and scalar field profile) within this theory had been found previously by Astorino in~\cite{Astorino:2013sfa}, and is given by:
\begin{eqnarray}
\label{eq:metriccc}
f_{\rm CC}(r)&=&1-\frac{2M}{r}+\frac{Q^2+S}{r^2}\,, \\
\label{bbmb2}
\phi&=&\pm \sqrt{\frac{6}{8\pi}}\sqrt{\frac{S}{Q^2+S}}\,.
\end{eqnarray}
where the continuous parameter $S$ characterizes the conformally-coupled scalar hair, and can be viewed as a scalar charge. As one can see in~\cite{Astorino:2013sfa}, the scalar charge $S$ emerges as an integration constant, characterizing primary hair. The BH solution with metric function given by Eq.~(\ref{eq:metriccc}) is commonly referred to in the literature as ``\textit{constant scalar hairy charged BH}'', reflecting the fact that the scalar field has a constant but non-zero value. We shall refer to this solution as conformally-coupled charged hairy black hole (CCCHBH). The CCCHBH solution is parametrized by two charges: the electric charge $Q$, and the scalar charge $S$. The fact that the scalar charge appears in the metric function with power unity (unlike the electric charge which appears quadratically) will play an important role in our subsequent discussion.~\footnote{It is interesting to mention that this solution is closely related to the so-called ``BBMB black hole''~\cite{Bocharova:1970skc,Bekenstein:1974sf}. However, the presence of the electromagnetic field improves the properties of the full solution, since the scalar field in the BBMB BH is non-regular. Furthermore, it was recently shown in~\cite{Zou:2019ays} that the CCCHBH solution in Eq.~\ref{bbmb2} is stable against perturbations.}

The geometry of the CCCHBH solution is very similar to that of the Reissner-Nordstr\"{o}m charged BH, since the positions of Cauchy and event horizons are given by (recall that we are working in units of CCCHBH mass) $r_{c,e}=1 \mp \sqrt{1-Q^2-S}$. If we only consider positive values for $S$, then the CCCHBH solution effectively describes a non-extremal RN BH since $0<S<1\,,0<Q<\sqrt{1-S}$. One can view this solution as describing a RN BH with effective charge $Q_{\rm eff} = \sqrt{Q^2+S}$. Equivalently, one can view the CCCHBH solution as a RN BH with extremal limit given by $Q_{\rm ext} = \sqrt{1-S}$, such that $Q_{\rm ext}<1$ for $S>0$.

However, there are two other possible regimes. In particular, one might allow for negative scalar hair, $S<0$. In this case, one has to consider the following two situations: $-Q^2<S<0$ or $S<-Q^2$. The first range can actually be discarded since, as shown in~\cite{Chowdhury:2018izv}, the kinetic term of the resulting action is trivial. More interesting is the regime where $S<-Q^2$. In this case, one can set the coefficient in the $1/r^2$ term of the CCCHBH metric function to be negative. As discussed in a number of works, and most recently in~\cite{Chowdhury:2018izv}, the resulting so-called \textit{mutated} Reissner-Nordstr\"{o}m BH can effectively mimic a so-called Einstein-Rosen bridge, or wormhole~\cite{Einstein:1935tc,Morris:1988tu}, a tunnel-like structure connecting two disparate points of space-time.~\footnote{Wormhole solutions usually need to be sustained by an exotic matter component which violates the null energy condition~\cite{Morris:1988tu}. An alternative possibility is that the null energy condition might be effectively violated in a controlled way in theories of gravity beyond General Relativity. This is achieved without the need to introduce exotic matter components, thanks to the effective stress-energy tensor due to the beyond-GR terms. This possibility has been studied widely especially in recent years: for an incomplete list of recent important works in this direction, see e.g.~\cite{Richarte:2007zz,Eiroa:2008hv,Richarte:2010bd,Bolokhov:2012kn,Capozziello:2012hr,
Harko:2013yb,Richarte:2013fek,DiCriscienzo:2013ria,Shaikh:2015oha,Myrzakulov:2015kda,Bahamonde:2016jqq,Ovgun:2016ujt,Sebastiani:2016ras,Moraes:2016akv,Ovgun:2017zao,Shaikh:2017zfl,Ovgun:2017jip,Jusufi:2017vta,Moraes:2017mir,Jusufi:2017mav,Moraes:2017dbs,Calza:2018ohl,Jusufi:2018kmk,Shaikh:2018kfv,Ovgun:2018xys,Amir:2018pcu,Shaikh:2018yku,Javed:2019qyg,Antoniou:2019awm,Gorji:2019rlm,Singh:2020rai}.} We remark that the mutated RN solution we are dealing with only mimics a wormhole, and we leave the question of whether the effective wormhole solution resulting within the regime $S<-Q^2$ is traversable~\cite{Visser:1989kh} to a future study. Instead, in the following we will be interested in understanding the effect of the scalar charge $S$ on the BH shadow. Therefore, for the case of the CCCHBH solution, the parameter $S$ will play the role of hair parameter. This is analogous to how the parameter $\nu$ played the role of hair parameter for the MCCHBH solution studied in Sec.~\ref{sec:mcchbh}. Finally, we note that since within this regime the radial coordinate of the Cauchy horizon becomes $r_c<0$, there is no Cauchy horizon, but only an event horizon.

\section{Shadows of conformally-coupled charged black holes with scalar hair}
\label{sec:shadowccchbh}

\begin{figure}[!ht]
\begin{center}
\includegraphics[width=0.5\linewidth]{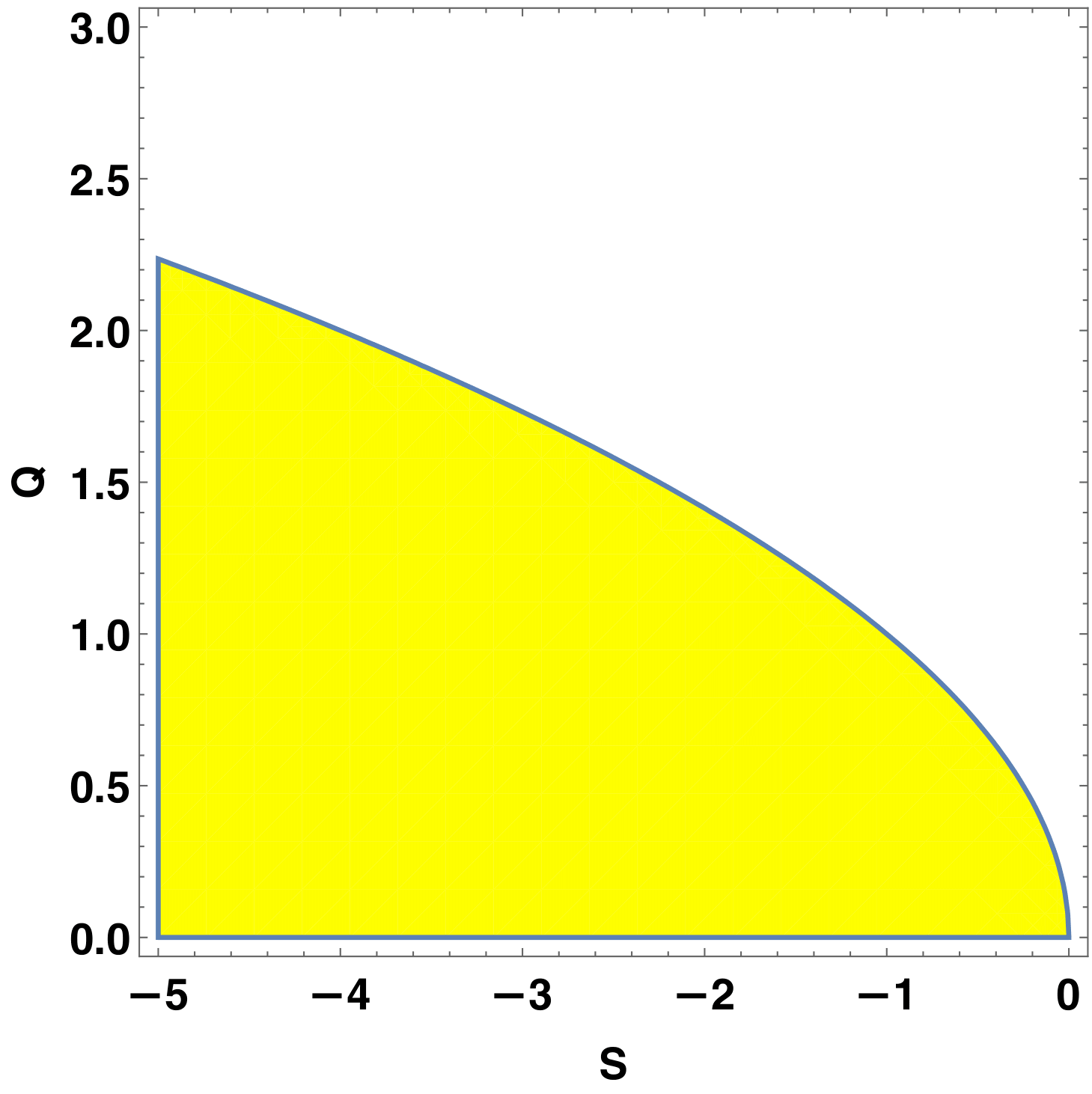}
\caption{Regime of validity of the CCCHBH solution given by Eqs.~(\ref{eq:metriccc},\ref{bbmb2}), obtained by imposing that the radial coordinates of all three the photon sphere $r_{ph}$, event horizon $r_e$, and Cauchy horizon $r_c$ be real-valued, within the regime $S<-Q^2$, with $S$ and $Q$ the scalar charge and electric charge respectively. Recall that in this regime the CCCHBH solution describes a mutated Reissner-Nordstr\"{o}m black hole, which effectively mimics a wormhole. The allowed area in the $S$-$Q$ parameter space is shaded in yellow.}
\label{C0}
\end{center}
\end{figure}

Having discussed the metric of the conformally-coupled charged black hole with scalar hair (CCCHBH), now want to compute the shadow it casts. To proceed, we can follow a completely analogous set of steps as in Sec.~\ref{sec:shadowmcchbh} to find the CCCHBH equivalent of Eqs.~(\ref{b},\ref{un}). Doing so we find:
\begin{eqnarray}
\label{bb}
&b^2=\frac{r^2}{f_{\rm CC}(r)}\,,\\
\label{unn}
&2f_{\rm CC}(r)-r\frac{df_{\rm CC}(r)}{dr}=0\,.
\end{eqnarray}
Explicitly plugging in the metric function given in Eq.~(\ref{bbmb2}), simple manipulations bring us to the following expression for the radial coordinate of the CCCHBH photon sphere:
\begin{eqnarray}
r_{ph}=\frac{3}{2} \left ( 1\pm\sqrt{1-\dfrac{8}{9} \left (Q^2+S \right ) } \right)\,,
\label{ph}
\end{eqnarray}
For the case where the BH carries no charge, neither electric nor scalar, \textit{i.e.} $Q=S=0$, we know that we should recover the Schwarzschild limit $r_{ph}=3$. From this we infer that we should only consider the positive sign in Eq.~(\ref{ph}). In the case where the scalar charge is positive, $S>0$, it is easy to show that $r_{ph}$ and $r_{c,e}$ are real-valued if and only if $0<Q<1$ and $0<S\leq 1-Q^2$, with the resulting solution describing a non-extremal RN BH. From our earlier discussion, we can also consider the regime $S<-Q^2$, where the solution describes a mutated RN BH, effectively mimicking a wormhole. In this regime, demanding that $r_{ph}$ and $r_{c,e}$ be real-valued leads to non-trivial conditions. We show this in Fig.~\ref{C0}, where the yellow-shaded region in the $Q$-$S$ plane is that allowed by the condition that $r_{ph}$ and $r_{c,e}$ be real-valued. As is clear from the figure, within this regime values of the electric charge $Q>1$, which would not be allowed for a standard RN BH (the extremal RN BH has $Q=1$), are now allowed if the scalar charge is sufficiently negative. We are not aware of any theoretical argument which sets a lower limit to the (negative) amount of scalar hair $S$.
\begin{figure}[!ht]
\begin{center}
\includegraphics[width=0.45\linewidth]{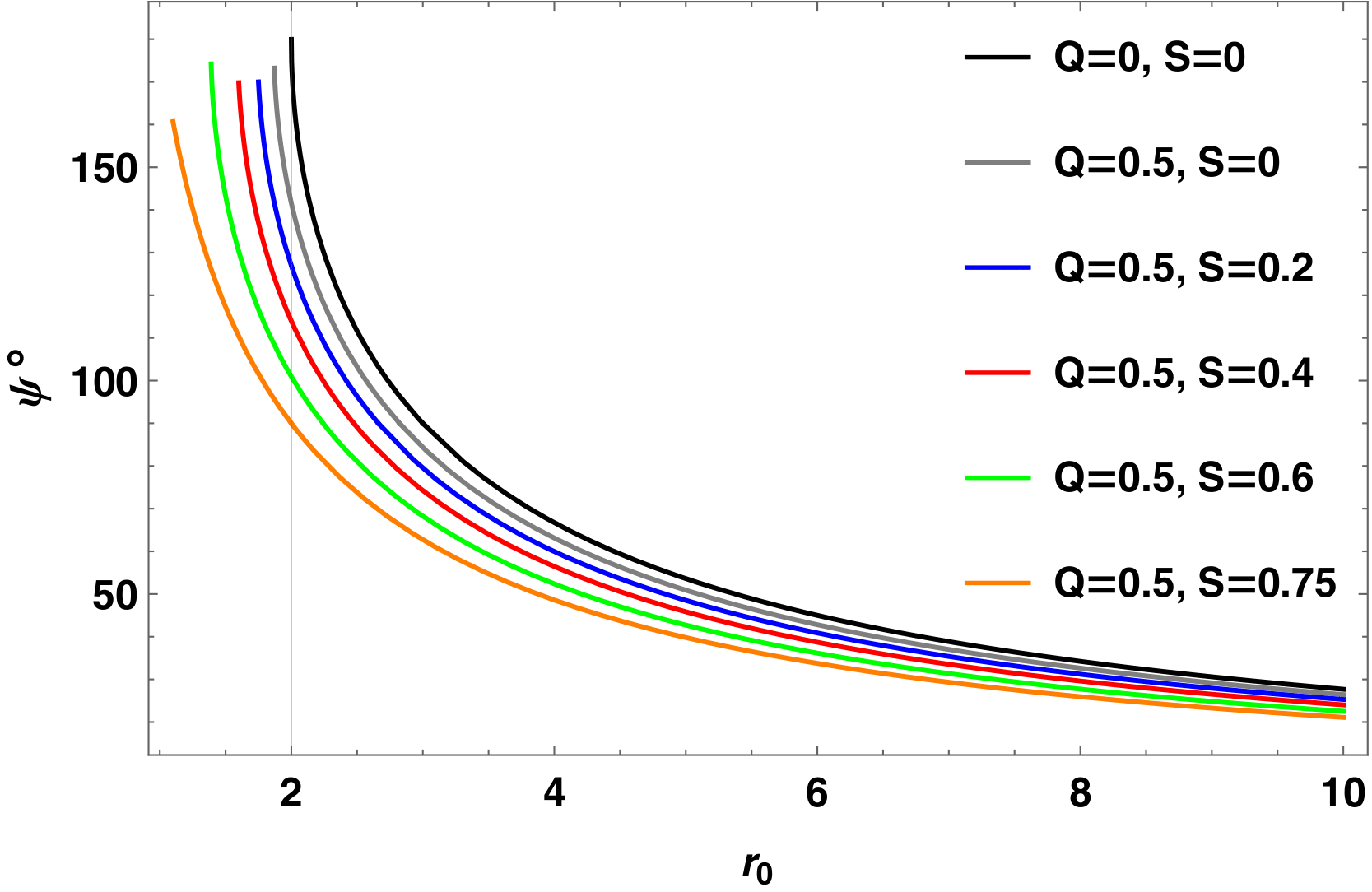}
\includegraphics[width=0.45\linewidth]{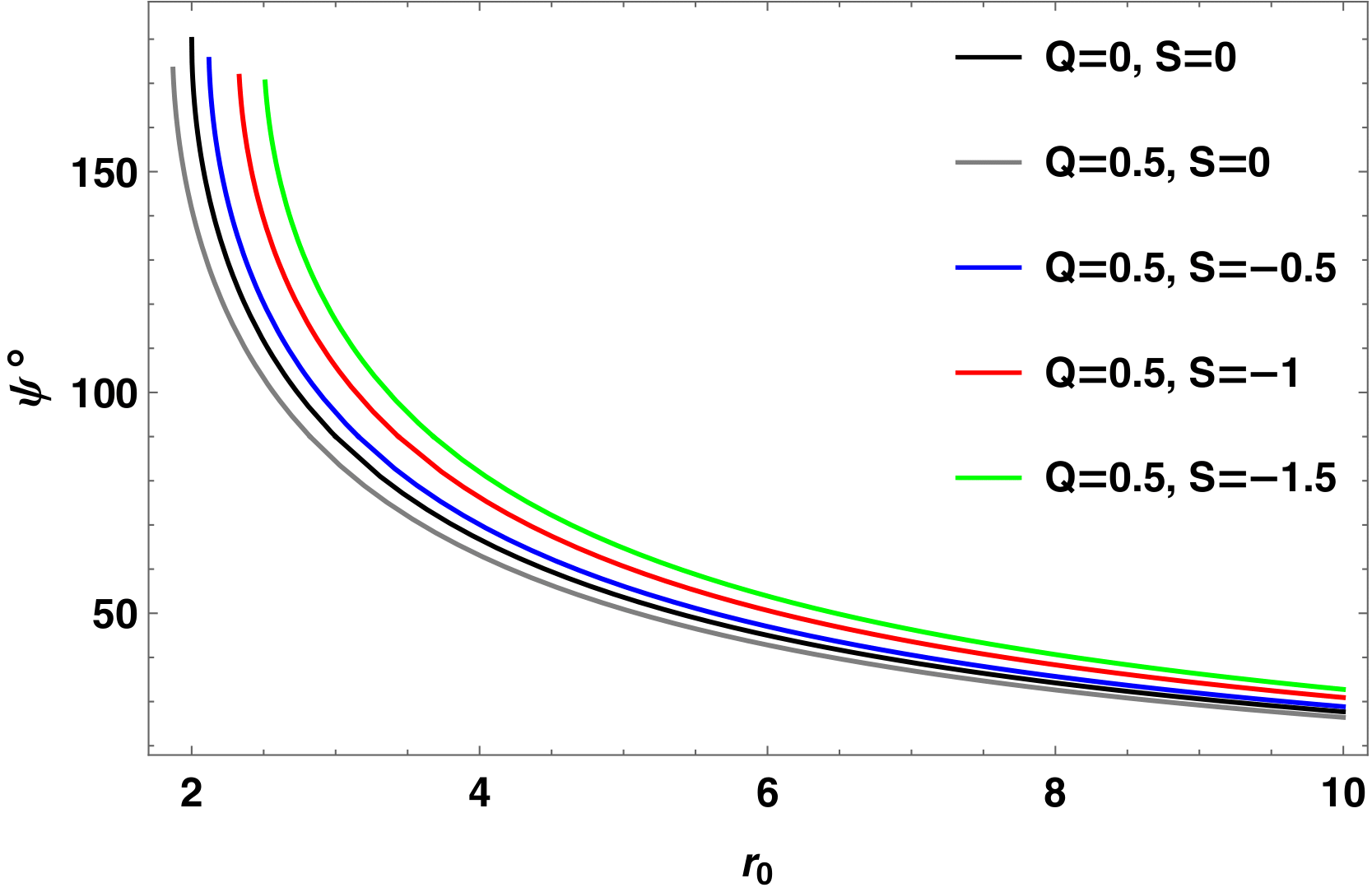}
\includegraphics[width=0.45\linewidth]{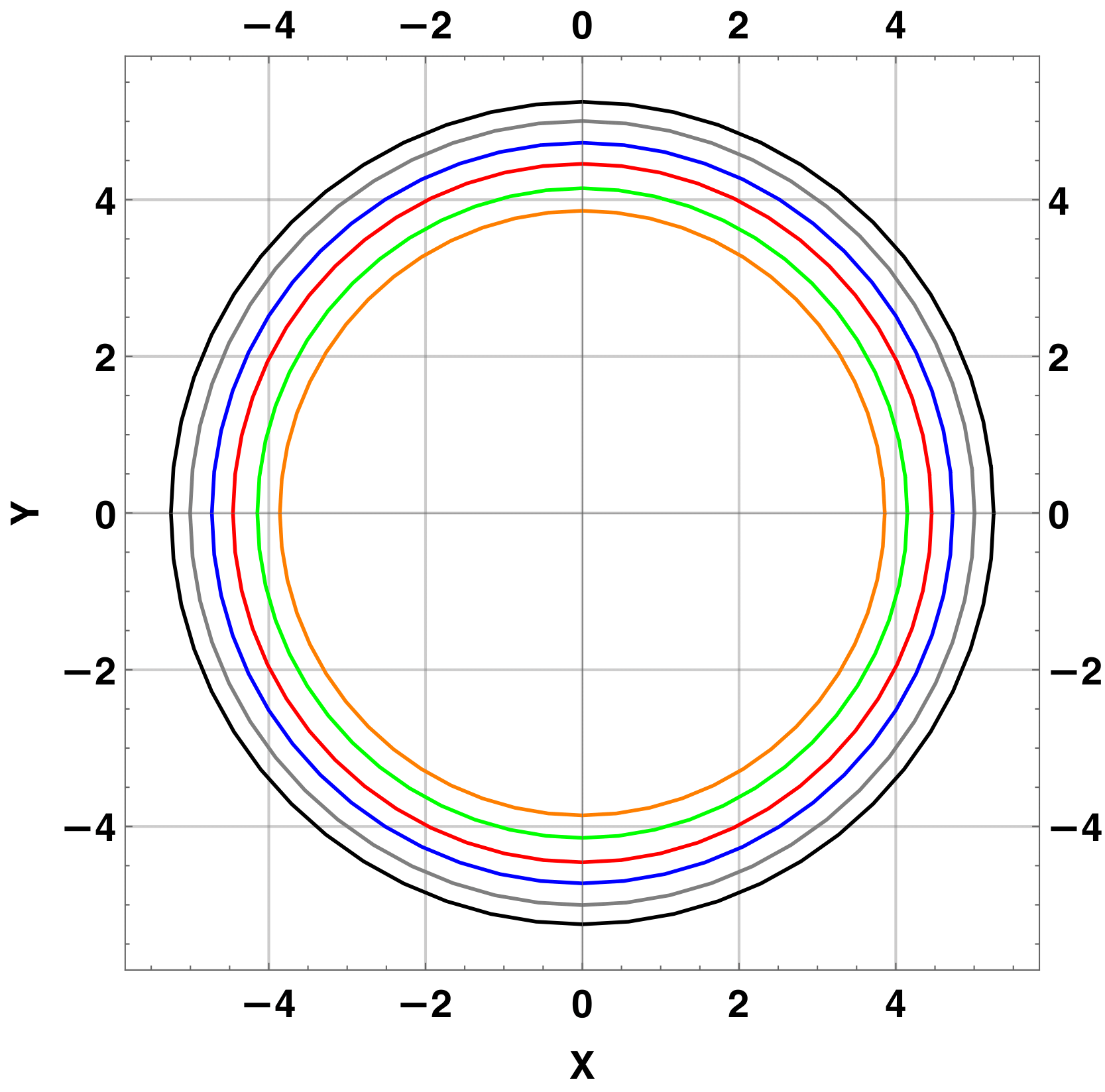}
\includegraphics[width=0.45\linewidth]{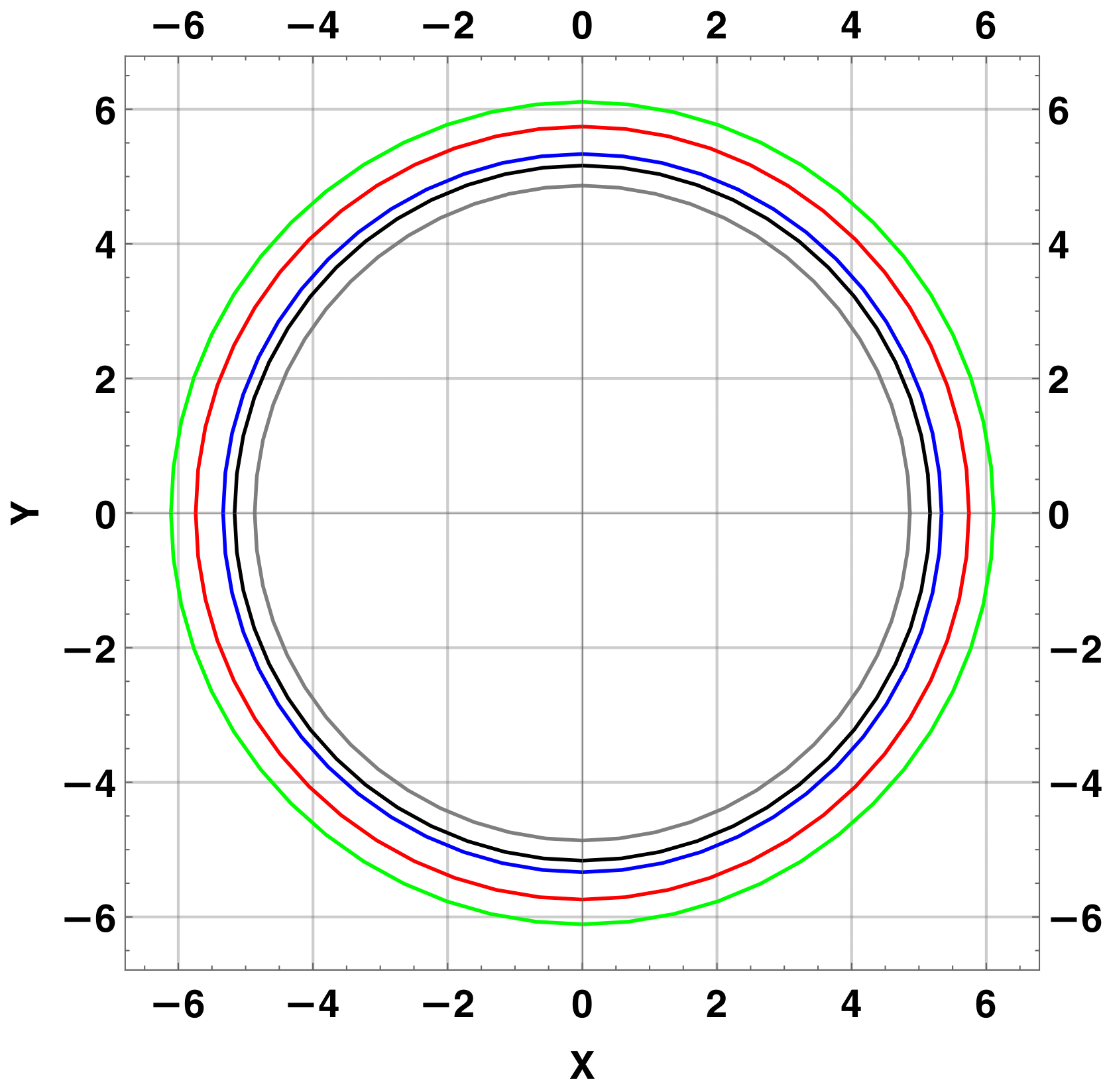}
\caption{\textit{Upper panels}: plot of the angle between the four-vector tangent to the path of a photon on the photon sphere and the position coordinate of a static distant observer, $\psi$, versus the radial coordinate of a distant observer $r$, for several values of the electric charge $Q$ and scalar charge $S$ of the conformally-coupled charged hairy black holes (CCCHBHs) studied in this section. The left and right panel consider values of $Q$ and $S$ within the two regimes $S>0$ and $S<-Q^2$ respectively. Recall that in the latter regime the CCCHBH solution describes a mutated Reissner-Nordstr\"{o}m black hole, which effectively mimics a wormhole. \textit{Lower panels}: shadows cast by the CCCHBHs, as seen by a static distant observer. The underlying values of the electric and scalar charges $Q$ and $S$ follow the same colour-coding as the upper panels, with the left and right panels considering the regimes $S>0$ and $S<-Q^2$ respectively.}
\label{CSH}
\end{center}
\end{figure}

As we did in Sec.~\ref{sec:shadowmcchbh}, we now compute $\psi$, the angle subtended between the four-vector tangent to the photon's path and the four-vector of a static distant observer at $r=r_0$. We show the result in the upper panels of Fig.~\ref{CSH}, for selected values of $Q$ and $S$, considering both positive (left panel) and negative (right panel) values of $S$. For the case where $S>0$, we see that $\psi$ decreases as $S$ is increased, meaning that increasing the amount of BH hair reduces the shadow size. This is consistent with the behaviour of hairless RN BHs, where increasing the amount of electric charge decreases the BH shadow size. For the case where $S<-Q^2$, we see that increasing the magnitude of $S$ (more negative $S$) increases $\psi$ and hence the shadow size. Therefore, the mutated RN BH/effective wormhole has a shadow which is larger than that of the corresponding Schwarzschild BH, something which cannot occur for the standard RN BH. Therefore, we expect the limits on the amount of scalar charge $S$ from the EHT shadow of M87* should be qualitatively different for the two regimes $S>0$ and $S<-Q^2$.

\section{Constraints on the amount of scalar hair from the Event Horizon Telescope shadow of M87*}
\label{sec:ehtshadow}

Let us recall what we have done so far. In Sec.~\ref{sec:shadowmcchbh} and Sec.~\ref{sec:ccchbh}, we have computed the shadows of the BHs resulting from two theories of GR plus a scalar field, in the first case minimally-coupled and in the second case conformally-coupled. The resulting BH solutions, denoted MCCHBH and CCCHBH respectively, were found to carry primary scalar hair. For the MCCHBH, the relevant hair parameter is associated to an explicit scale introduced in the scalar field profile and denoted by $\nu$. For the CCCHBH, the relevant hair parameter is the scalar charge $S$. Both BHs also carry electric charge $Q$. In this section, we shall set joint limits on the electric charge and the hair parameter of these two clases of BHs, using the Event Horizon Telescope (EHT) image of the dark shadow of M87*. As we have already shown in Fig.~\ref{SH1} and Fig.~\ref{CSH}, the size of the resulting shadow can depend rather strongly on the amount of scalar hair. We can therefore expect the image of M87* to set meaningful constraints on the hair parameters $\nu$ and $S$.

The EHT collaboration reports in~\cite{Akiyama:2019cqa,Akiyama:2019eap} that the angular size of the shadow of M87* is $\delta = (42 \pm 3)\,\mu{\rm as}$ and the distance to M87* is $D = 16.8^{+0.8}_{-0.7}\,{\rm Mpc}$. Finally, associating the crescent feature in the image of M87* with the emission surrounding the BH shadow, and further knowing the distance to M87*, the EHT collaboration estimated the mass of M87* to be $M = (6.5 \pm 0.9) \times 10^9\,M_{\odot}$~\cite{Akiyama:2019eap}. We can combine these numbers into the single number $d_{M87*}$ which measures the angular size of M87*'s shadow in units of mass, and therefore enables a direct comparison to the shadows we already computed. In particular, $d_{M87*}$ is given by:
\begin{eqnarray}
d_{M87*} \equiv \frac{D\delta}{M} \approx 11.0 \pm 1.5\,.
\label{size}
\end{eqnarray}
The detected angular size of M87*'s shadow in units of its mass, as given in Eq.~(\ref{size}), is remarkably consistent with that of the Schwarzschild BH (recall for a Schwarzschild BH $d=6\sqrt{3} \simeq 10.4$. Within $1\sigma$ uncertainties, Eq.~(\ref{size}) gives $9.5 \lesssim d_{M87*} \lesssim 12.5$, which easily encompasses the Schwarzschild value $6\sqrt{3}$.

\begin{figure}[!ht]
\begin{center}
\includegraphics[width=0.5\linewidth,height=4.5cm]{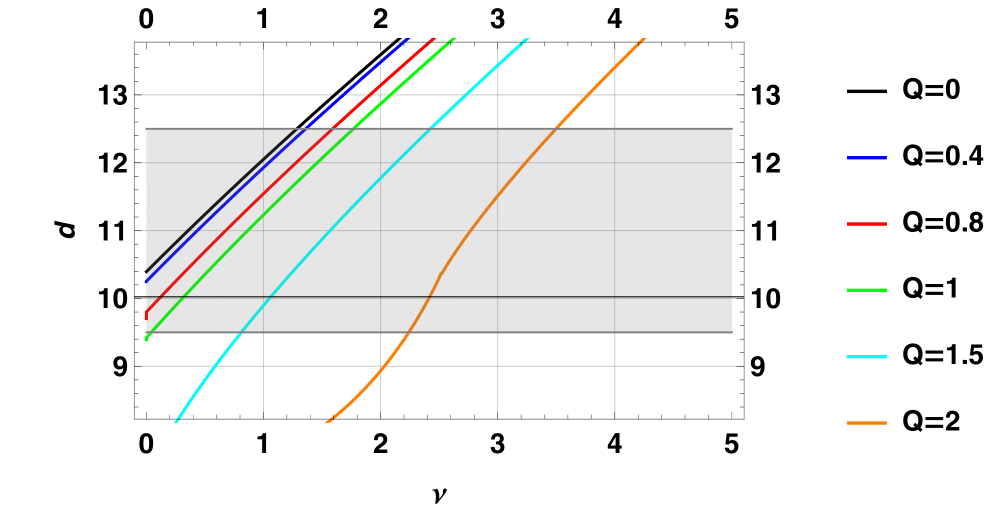}
\includegraphics[width=0.4\linewidth]{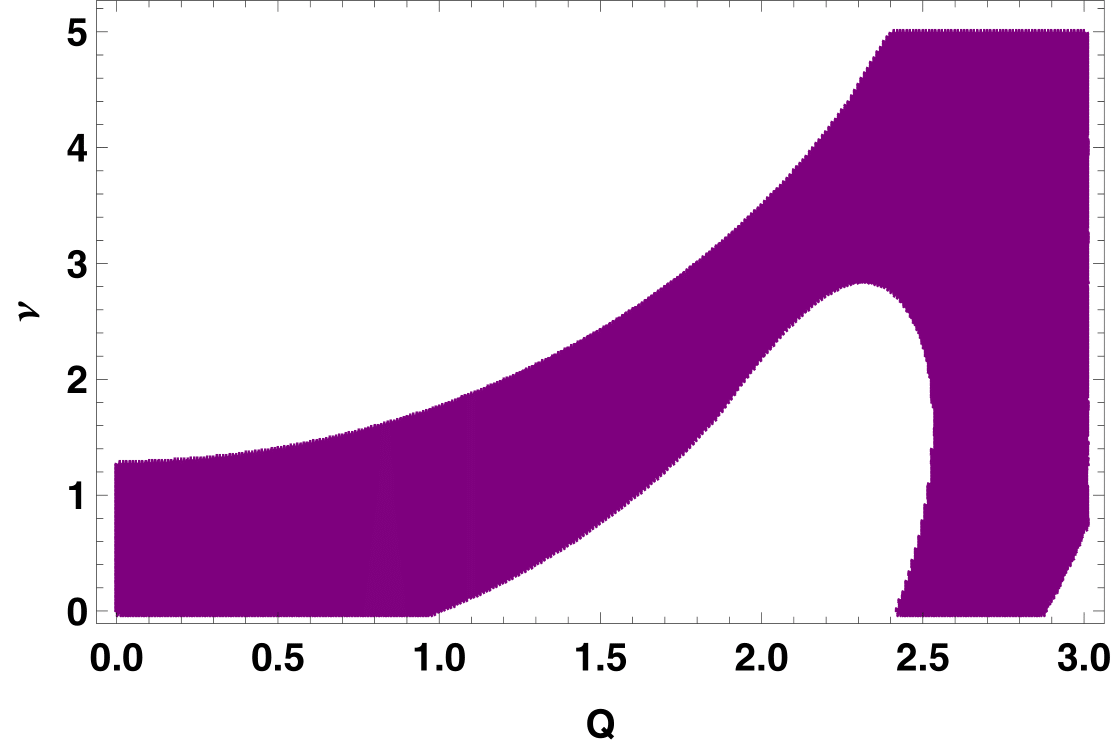}
\caption{\textit{Left panel}: diameter of the shadow of the minimally-coupled charged black holes with scalar hair (MCCHBH) computed in Sec.~\ref{sec:shadowmcchbh} in units of mass $M$, as a function of the hair parameter $\nu$, for various values of the MCCHBH electric charge $Q$. The grey shaded region indicates the $1\sigma$ credible region for the diameter of the shadow of the supermassive BH M87* imaged by the Event Horizon Telescope, as given by Eq.~(\ref{size}). \textit{Right panel}: the purple-shaded region is the region of $Q$-$\nu$ parameter space where the diameter of the resulting MCCHBH shadow is compatible with the shadow of M87* within $1\sigma$. If we restrict our attention to the observationally viable region for which $Q \ll M$, we see that the EHT sets a rough upper limit of $\nu \lesssim {\cal O}(M)$ on the amount of scalar hair carried by the MCCHBH.}
\label{PC0}
\end{center}
\end{figure}

We begin by considering the MCCHBH solution whose shadow we computed in Sec.~\ref{sec:shadowmcchbh}, and with hair parameter given by $\nu$ as introduced in Eq.~(\ref{field}). In Fig.~\ref{PC0}, we plot the diameter of the resulting BH shadow as a function of the hair parameter $\nu$ for selected values of $Q$, together with the $1\sigma$ credible region on $d_{M87*}$ from the EHT detection (left panel). The results reflect the behaviour we had already observed in Fig.~\ref{SH1}. Subsequently, we scan the $Q$-$\nu$ parameter space and select the region where the diameter of the resulting shadow respects the limit set by the EHT detection of the shadow of M87* as given in Eq.~(\ref{size}). The resulting allowed region of parameter space is shaded in purple in the right panel of Fig.~\ref{PC0}.

As we see from the allowed region of parameter space shown in the right panel of Fig.~\ref{PC0}, there are qualitatively two different situations depending on the size of $Q$. For $0<Q<1$, the standard region allowed for a RN BH, we find a rough upper limit of $\nu \lesssim 1$ (recall this means $\nu \lesssim M$ as the hair parameter $\nu$ carries dimensions of mass). Then, as we increase $Q$ beyond the RN limit $Q=1$, we find both upper and lower limits on $\nu$, such that $1.2 \lesssim \nu \lesssim 1.6$ is required for the resulting shadow size to be consistent with that of M87*. Once $Q$ gets sufficiently large ($Q \gtrsim 2$), basically any value of $\nu$ becomes allowed. The result, as one can appreciate from the lower right panel of Fig.~\ref{SH1} and the values of $r_{ph}$ shown in our numerical examples in Tab.~\ref{Nu}, is due to the fact that for $Q \gtrsim 0.1$, $Q$ and $\nu$ essentially work in opposite directions: increasing the former would make the shadow size decrease (as we expect for standard RN BHs), whereas increasing the latter would make the shadow size increase. Therefore, for sufficiently large $Q$, one can in principle always find an equally large $\nu$ such that the resulting shadow size is compatible with that of M87*. Therefore, we expect their allowed regions of parameter space to be positively correlated.

At this point, it is important to point out that we expect realistic astrophysical BHs to carry a small amount of electric charge. That is, $Q \ll M$. This is a purely observational requirement arising from the fact that we have not observed BHs which carry a significant amount of electric charge, close to the RN limit $Q=1$. If we therefore restrict our attention to the observationally viable region of parameter space where $Q \ll 1$, we are able to set a rough upper limit of $\nu \lesssim {\cal O}(M)$ on the amount of scalar hair carried by the MCCHBH.

\begin{figure}[!ht]
\begin{center}
\includegraphics[width=0.47\linewidth]{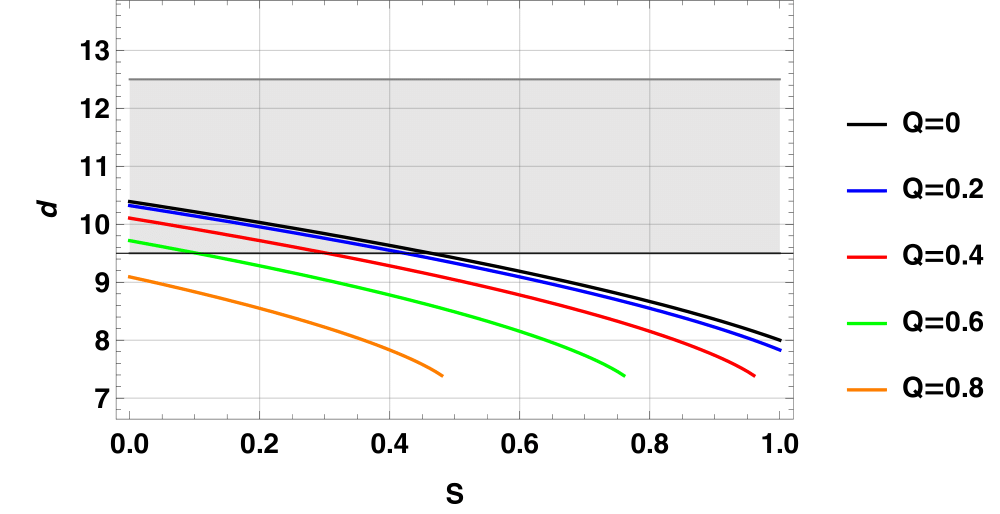}
\includegraphics[width=0.47\linewidth]{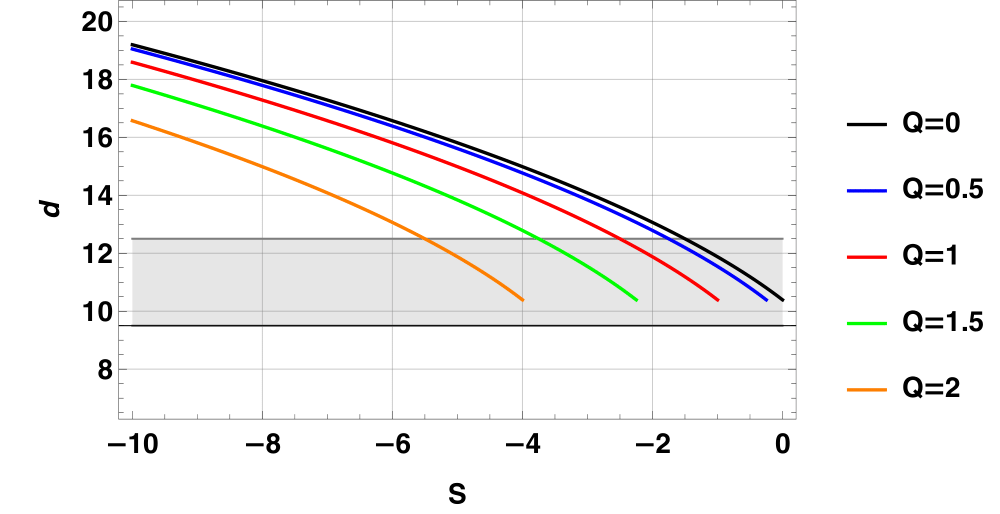}
\caption{Diameter of the shadow of the conformally-coupled charged black holes with scalar hair (CCCHBH) computed in Sec.~\ref{sec:shadowccchbh} in units of mass $M$, as a function of the hair parameter $S$, for various values of the MCCHBH electric charge $Q$. The grey shaded regions indicate the $1\sigma$ credible region for the diameter of the shadow of the supermassive BH M87* imaged by the Event Horizon Telescope, as given by Eq.~(\ref{size}). The left and right panel consider values of $Q$ and $S$ within the two regimes $S>0$ and $S<-Q^2$ respectively. Recall that in the latter regime the CCCHBH solution describes a mutated Reissner-Nordstr\"{o}m black hole, which effectively mimics a wormhole.}
\label{PC1}
\end{center}
\end{figure}

\begin{figure}[!ht]
\begin{center}
\includegraphics[width=0.45\linewidth]{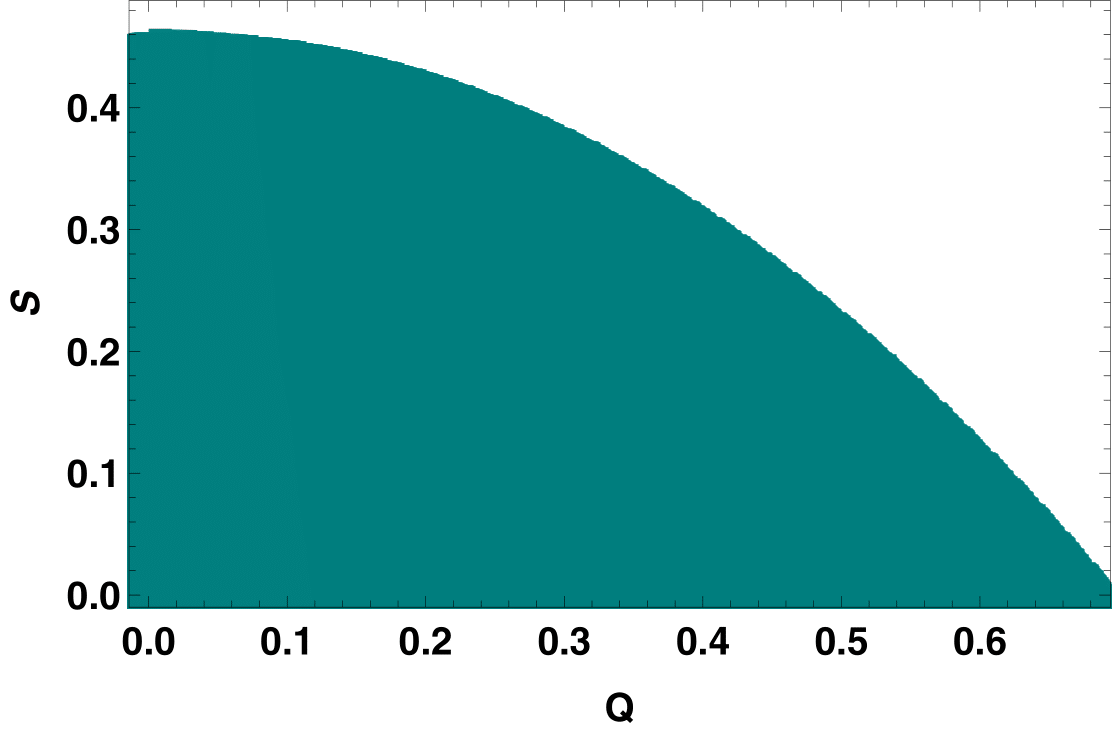}
\includegraphics[width=0.45\linewidth]{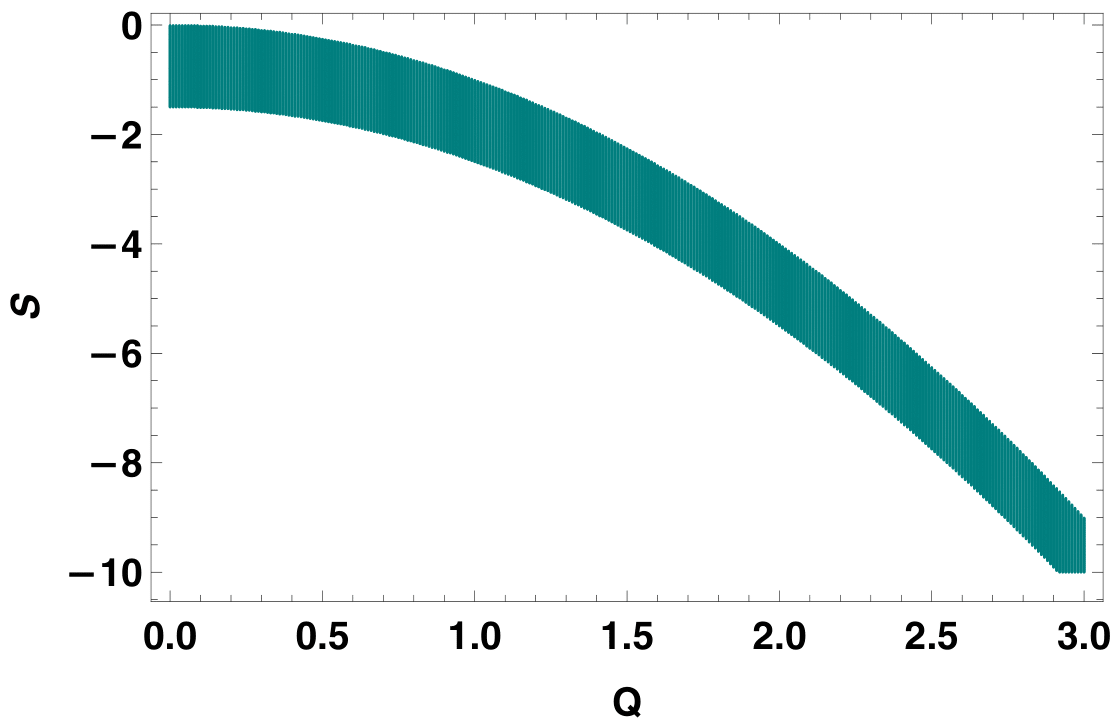}
\caption{The teal-shaded region is the region of $Q$-$S$ parameter space where the diameter of the resulting shadows of the conformally-coupled charged black holes with scalar hair (CCCHBHs) are compatible with the shadow of M87* within $1\sigma$, as given by Eq.~(\ref{size}). The left and right panels consider values of $Q$ and $S$ within the two regimes $S>0$ and $S<-Q^2$ respectively. Recall that in the latter regime the CCCHBH solution describes a mutated Reissner-Nordstr\"{o}m black hole, which effectively mimics a wormhole. If we restrict our attention to the observationally viable region for which $Q \ll 1$, we see that the EHT sets a rough upper limit of $S \lesssim 0.4M^2$ within the regime $S>0$ and a rough lower limit of $S \gtrsim -1.5M^2$ within the regime $S<-Q^2$, on the amount of scalar hair carried by the CCCHBH.}
\label{PS}
\end{center}
\end{figure}

We now turn our attention to the CCCHBH solution whose shadow we computed in Sec.~\ref{sec:ccchbh}, and with hair parameter given by $S$ as introduced in Eq.~(\ref{eq:metriccc}). In Fig.~\ref{PC1}, we plot the diameter of the resulting BH shadow as a function of the hair parameter $S$ for selected values of $Q$, alongside the $1\sigma$ credible region on $d_{M87*}$ from the EHT detection, for the two regimes we have discussed earlier: $S>0$ (left panel), and $S<-Q^2$ which can described a mutated RN BH/wormhole (right panel). Subsequently, we scan the $Q$-$S$ parameter space and select the region of parameter space where the diameter of the resulting shadow respects the limit set by the EHT detection of the shadow of M87* as given in Eq.~(\ref{size}). The resulting allowed region of parameter is shaded in teal in Fig.~\ref{PS}, for the two regimes we have discussed earlier: $S>0$ (left panel) and $S<-Q^2$ (right panel).

As we could have expected given our earlier separate discussions about the two regimes $S>0$ and $S<-Q^2$, the limits in $Q$-$S$ parameter space we obtain are qualitatively different within these two regimes. In fact, as we see from Fig.~\ref{PC1}, increasing the absolute value of the hair parameter $S$ leads to the shadow size decreasing for $S>0$ and increasing for $S<-Q^2$. For the case where $S>0$, which requires $Q<\sqrt{1-S}$ we see that we can place upper limits on the hair parameter $S$ of order $S \lesssim {\cal O}(0.1)$ (recall this means $S \lesssim {\cal O}(0.1)M^2$ as the hair parameter $S$ carries dimensions of mass squared) for $Q \lesssim 0.6$, with the upper limit getting tighter as $Q$ is increased (for $Q=0$, one gets $S \lesssim 0.4$). For $Q \gtrsim 0.7$, the upper limit on $S$ drops to $0$, meaning that for larger values of the electric charge, no amount of positive scalar hair/charge is allowed by the EHT detection of the shadow of M87*. In other words, not only is a positive amount of scalar charge excluded at $1\sigma$, but more generally this exclusion extends to RN BHs with $Q \gtrsim 0.7$. As we see in Fig.~\ref{PC1} (left panel), this is due to the fact that increasing $S$ within this regime decreases the shadow size more than what is allowed by the EHT observation. For instance, we see that the $Q=0.8$ case (orange curve in the left panel of Fig.~\ref{PC1}) is completely outside the $1\sigma$ EHT-allowed grey region, regardless of the value of $S$, and hence is excluded.

However, if we restrict our attention to the observationally viable region of parameter space for which $Q \ll M$, due to the lack of observation of BHs carrying a significant amount of electric charge, then we see that we can set a rough upper limit of $S \lesssim 0.4M^2$ on the amount of scalar hair carried by the CCCHBH assuming $S>0$.

Within the regime $S<-Q^2$, where the solution carries a negative amount of scalar hair/charge, and values of $Q$ beyond the RN limit $Q=1$ are allowed, we see a qualitatively different behaviour. Here we find that, for any value of $Q$, there is a restricted range of values of $S$ for which the resulting shadow has a size consistent with that of M87*. The reason is that, within this regime, $Q$ and $S$ operate in opposite directions, and hence their allowed regions of parameter space are expected to be negatively correlated. In fact, within this regime, raising $Q$ acts to further decrease the shadow size, whereas lowering $S$ acts to increase the shadow size. Therefore, for any given value of $Q$, however extreme, one can always find an equally extreme range of values of $S$ for which the resulting shadow has a size compatible with the $1\sigma$ region allowed by the EHT observation of M87*: this behaviour works analogously, albeit in the opposite direction, compared to the positive $Q$-$\nu$ correlation we saw earlier for the MCCHBH. In this sense, the EHT observation is, at the level of angular size of the shadow, consistent with M87* being a wormhole described by an effective mutated Reissner-Nordstr\"{o}m metric arising from a theory with a scalar field conformally-coupled to GR.~\footnote{Recently, in an interesting study~\cite{Dai:2019mse}, a novel method for detecting wormholes via perturbations in the orbit of stars near supermassive BHs was proposed and applied in~\cite{Simonetti:2020vhw}. Another interesting method was proposed in~\cite{Dent:2020nfa}, making use of peculiar (anti-chirp) signatures imprinted in the waveform of gravitational waves produced by a BH orbiting a wormhole. See e.g.~\cite{Tsukamoto:2012xs,Tsukamoto:2016qro,Tsukamoto:2017edq} for further examples of observational tests of the existence of wormholes.} As an important caveat it is worth noting that this conclusion was reached by exclusively considering the shadow size, since we have not included any other observation. We therefore advise the reader to treat this intriguing conclusion with caution. Finally, if we once more restrict to the observationally viable region of parameter space where $Q \ll 1$, then we see that we can set a rough limit of $-1.5 \lesssim S/M^2 \lesssim 0$ on the amount of scalar hair carried by the CCCHBH assuming $S<-Q^2$.

In closing this section comparing the shadows we computed to that of M87*, three important comments and caveats on our results are in order. The first is that both the MCCHBH and CCCHBH solutions we have considered are static, \textit{i.e.} non-rotating. In other words, we have loosely speaking considered Reissner-Nordstr\"{o}m BHs with scalar hair. However, most of the BHs we have observed, including M87*, carry spin~\cite{Tamburini:2019vrf}. This spin can have a potentially important effect on the resulting shadow. In order to be fully consistent, we should therefore have incorporated rotation from the very beginning, studying Kerr-Newman BHs carrying scalar hair.

However, for standard BHs we know that the effect of angular momentum is to make the shadow asymmetric or oblate, since the effective potential on the side corresponding to photons with angular momentum aligned with the BH spin is higher, resulting in the BH shadow being flattened along the corresponding side. Such a deviation is small, resulting in a deviation from circularity $\Delta C$ which is $\gtrsim 10\%$ only for high observation angles, $\theta_{\rm obs} \simeq \pi/2$, when the BH is viewed edge-on. For M87*, it is believed that the mechanism powering the BH jet is the Blandford-Znajek mechanism, or a closely related one, wherein the jet is powered by the magnetic field of the accretion disk extracting energy from the BH spin. If the jet is powered by the BH spin, one expects the two to be roughly aligned. In this case, the observation angle of M87* should roughly equal that of the jet, $\theta_{\rm obs} \approx 17^{\circ}$, so that the BH is viewed nearly face-on. For such low observation angles, the effect of spin is expected to be very small in terms of oblateness. Hence the shadow sizes we have computed earlier and used to compare our results against the EHT observation should hold to very good approximation.

Of course, the previous statement is true for standard hairless BHs. While we expect this conclusion to qualitatively hold for hairy BHs as well, it is possible that the presence of scalar hair might amplify the effect of the BH spin on the shadow oblateness/size, or make this effect non-trivial. To completely address this issue, it would be interesting to explicitly construct the rotating versions of the MCCHBH and CCCHBH, for instance by making use of the Newman-Janis algorithm. Since doing so would considerably complicate our analysis and would be beyond the scope of the present work, we leave this issue for a follow-up project.

The second important caveat is that our constraints on the scalar hair parameters $\nu$ and $S$ are of course model-dependent, and only hold for the specific MCCHBH and CCCHBH solutions obtained within the specific (minimally- or conformally-coupled) scalar field theories. In principle, it could be desirable to find a more model-independent measure of the amount of BH hair. One interesting possibility in this sense is that put forward in~\cite{Cunha:2019ikd}, where the amount of (ultralight) scalar hair is measured via a parameter $p=1-M_H/M$, where $M_H$ is the Komar mass (or horizon energy), and $M$ is the ADM mass. The hair parameter $p$ is bounded within the range $0 \leq p \leq 1$, with larger values of $p$ indicating a larger amount of hair. The work of~\cite{Cunha:2019ikd} finds that the EHT observation of the shadow of M87* is consistent with at most $p \approx 10\%$. It would be interesting to make contact with this measure of the amount of hair, and we again leave this issue for a follow-up project.

The final caveat is related to our identification of the dark region in the image of M87* as the apparent image of the photon sphere. One might legitimately worry that the size of the shadow might be sensitive to the accretion flow details, making the size of the dark region unsuitable for comparison against theoretically computed shadows: a concern along these lines was recently raised in~\cite{Gralla:2019xty}. While this concern most certainly holds for the bright ring surrounding the dark region (whose size, morphology, and luminosity all depend very sensitively on the details of the accretion flow), it is known that for advection dominated accretion flow (ADAF) the dark region in VLBI-detected images is indeed the apparent image of the photon sphere: this was recently confirmed in~\cite{Narayan:2019imo}. The ADAF model is believed to accurately describe the accretion flow around low-redshift supermassive BHs, including M87* and SgrA*~\cite{Yuan:2014gma}. In this case, one can truly hope to use the morphology (and in particular the size) of the dark region in the image of M87* to learn about fundamental physics. We thus believe that our approach of comparing the size of the dark region imaged by the EHT against the angular size of the shadows we computed theoretically is relatively robust against details of the accretion flow.

\section{Conclusion}
\label{sec:conclusions}

The apparent simplicity of black holes, as exemplified by the no-hair theorem, will be thoroughly tested in the coming decade. These tests will be enabled owing to exquisitely precise probes of gravity in the strong-field regime, including gravitational waves and VLBI black hole shadow imaging. At the same time, advances in our understanding of the AdS/CFT correspondence call for more detailed studies on the behaviour of charged black holes in the presence of matter fields. We therefore believe that the time is ripe, from both the observational and theoretical points of view, to study controlled violations of the no-hair theorem, and observational signatures thereof. This is the big picture goal of our work.

We have considered two physical systems, where the gravitational sector is described by General Relativity, supplemented by the presence of an electromagnetic field. We have then added a matter source in the form of a scalar field, in the first case minimally-coupled, and in the second case conformally-coupled to gravity. We have then studied BH solutions within these two systems, finding that both carry primary scalar hair: we have referred to the two BH solutions as MCCHBH (minimally-coupled charged hairy black hole) and CCCHBH (conformally-coupled charged hairy black hole) respectively. The hair parameters characterizing the MCCHBH and CCCHBH were denoted by $\nu$ and $S$ respectively. It is worth noting that the CCCHBH admits a regime where the scalar charge satisfies $S<-Q^2$, with $Q$ the BH's electric charge. The resulting mutated Reissner-Nordstr\"{o}m BH was earlier found to effectively mimic a wormhole~\cite{Chowdhury:2018izv}. Subsequently, by studying the motion of photons in the vicinity of these two charged hairy BHs, we have computed the shadows they cast as a function of their electric charge and hair parameter.

Finally, we have proceeded to confront these shadows with the shadow of the supermassive BH M87*, as recently detected by the Event Horizon Telescope collaboration. We have exploited the fact that the angular size of M87*'s shadow is remarkably consistent with that of a Schwarzschild BH (once its mass is known) to set limits on the parameters of the two charged hairy BHs we have considered: electric charge $Q$, and hair parameter $\nu$ or $S$. For the case where the scalar field is minimally-coupled, the results are given in the right panel of Fig.~\ref{PC0} ($1\sigma$ confidence region). We find that a large region of $Q$-$\nu$ parameter space is allowed by the EHT observation. However, if we restrict our attention to the observationally preferred region where $Q \ll M$ (with $M$ the BH mass, given the lack of observation of BHs carrying significant electric charge), we are able to set a bound of $\nu \lesssim {\cal O}(M)$.

We have then repeated the analysis for the case where the scalar field is conformally-coupled, studying the two regimes $S>0$ and $S<-Q^2$ separately. The results are given in Fig.~\ref{PS}, with the two different regimes corresponding to the left and right panels respectively ($1\sigma$ confidence regions). If we again restrict to the observationally preferred region where $Q \ll M$, we see that we can set an upper limit of $S \lesssim 0.4M^2$ within the first regime, and a lower limit of $S \gtrsim -1.5M^2$ within the second regime. It is therefore interesting to note that the Event Horizon Telescope observation of M87* is in principle consistent with the object in question being a mutated Reissner-Nordstr\"{o}m black hole, which effectively mimics a wormhole. As an important caveat it is worth noting that this conclusion was reached by exclusively considering the shadow size, since we have not included any other observation. We therefore advise the reader to treat this intriguing conclusion with caution.

There are a huge number of avenues for follow-up work, including tying up a few loose ends and caveats which we have discussed at the end of Section~\ref{sec:ehtshadow}. For example, it would be interesting to extend our study to include rotation, therefore focusing on Kerr-Newman black holes with scalar hair. Moreover, the intriguing mutated Reissner-Nordstr\"{o}m regime in the conformally-coupled case is definitely worth further study. If M87* truly is a wormhole, it would be interesting to look for orthogonal signatures which could confirm this nature, including the motion of test particles in its vicinity. Finally, beyond the two well-motivated but still specific models we have studied in this work, it would be interesting to use the shadow to M87* to provide more general model-agnostic tests of the no-hair theorem, for instance characterizing the amount of scalar hair as done in~\cite{Cunha:2019ikd}. We leave these very interesting issues to future work. Overall, our work is among the first to provide concrete constraints on new physics beyond the Standard Model from the Event Horizon Telescope image of M87*: in particular, we are among the first, together with~\cite{Cunha:2019ikd}, to provide concrete tests of the no-hair theorem. The time is ripe to be working at the interface of theory and observations in the quest towards understanding gravity in the strong-field regime, and we believe our work represents a small but concrete step in this direction.

\acknowledgments

M.Kh. and A.A. are grateful to Hassan Firouzjahi for his supports. S.V. acknowledges support from the Isaac Newton Trust and the Kavli Foundation through a Newton-Kavli Fellowship, and acknowledges a College Research Associateship at Homerton College, University of Cambridge. D.F.M. acknowledges support from the Research Council of Norway.

\bibliographystyle{JHEP}
\bibliography{BH}

\end{document}